\documentclass[longbibliography,aps,prx,amsmath,amssymb,amsfonts,twocolumn,superscriptaddress,footinbib,floatfix]{revtex4-2}

\usepackage{amsmath}
\usepackage{amssymb}
\usepackage{amsfonts}
\usepackage{graphicx}
\usepackage{dcolumn}
\usepackage{bm}
\usepackage{hyperref}
\usepackage{xurl}
\usepackage{braket}
\graphicspath{{figures/}}
\hypersetup{hypertexnames=false,colorlinks=true,linkcolor=blue,citecolor=blue,urlcolor=blue}
\newcounter{proposition}
\newcommand{\mainstatement}[3]{\refstepcounter{proposition}\phantomsection\par\medskip\noindent\textbf{#1~\theproposition\ (#2).}\label{#3}\ }

\begin{document}

\title{A Finite-Window Recovery Hierarchy for Local Quantum Memory}

\author{Zheng An}
\email{anzheng@quantumsc.cn}
\affiliation{Quantum Science Center of Guangdong-Hong Kong-Macao Greater Bay Area (Guangdong), Shenzhen 518045, China}
\author{Dongyang Cao}
\affiliation{Quantum Science Center of Guangdong-Hong Kong-Macao Greater Bay Area (Guangdong), Shenzhen 518045, China}
\author{Jiangyu Cui}
\email{cuijy@quantumsc.cn}
\affiliation{Quantum Science Center of Guangdong-Hong Kong-Macao Greater Bay Area (Guangdong), Shenzhen 518045, China}
\date{\today}

\begin{abstract}
When quantum information initially stored in a local qubit disappears from that qubit, it need not be lost: it may have moved into nearby degrees of freedom, or it may remain there in a form that is not accessible to shallow local control. In this study, we introduce finite-window recoverability as an operational channel benchmark that separates these possibilities. The benchmark compares optimal recovery from the target site, recovery achieved by a bounded-depth decoder acting on a finite window, and the unrestricted optimum for that same window. Its operational component, local variational recovery, uses local state preparation, window-local control, and target-qubit Pauli readout to certify recoverable memory beyond the target site and to quantify how much of the same-window advantage is accessible to shallow control. In a disordered kicked-Ising Floquet chain, a depth-6 decoder on a five-site window realizes the hierarchy $Q^{\mathrm{opt}}_0<Q^{\mathrm{shallow}}_2<Q^{\mathrm{opt}}_2$ across the crossover regime, with positive certified gain for most disorder realizations and substantial shallow-accessibility fractions. The signal is distinct from target-site persistence and from reconstructed coherent-information increments, and the positive radius-2 gain persists when the same local task is embedded in longer open chains using an independent tensor-network backend. Guided by the hierarchy, we further test a carrier-deletion task in which the original target register is reset after the dynamics. A depth-8 decoder repairs the input from a radius-3 surrounding halo with held-out median $F_{\mathrm{avg}}=0.758$, above the single-qubit classical benchmark $2/3$, outperforming optimal one-, two-, and three-site halo-subwindow counterfactuals. These results establish finite-window recovery as a local-control benchmark for off-site quantum memory: it diagnoses not only where local quantum information remains, but whether it can be refocused by bounded-depth control.
\end{abstract}

\maketitle
\raggedbottom
\section{Introduction}
When local quantum information ceases to be recoverable from a qubit, the loss of on-site recoverability does not by itself determine whether that information has been destroyed or displaced into a nearby neighborhood. Long-lived local memory is a central aspect of the many-body-localization problem in disordered interacting systems~\cite{PhysRevLett.95.206603,Basko2006}. Interacting spin-chain studies established this phenomenology in a form close to the present setting~\cite{PhysRevB.77.064426,PhysRevB.82.174411}. Reviews of the subject emphasize that memory retention, transport, and thermalization need not be diagnosed by the same quantities~\cite{Nandkishore2015,Alet2018,RevModPhys.91.021001}. The same distinction is natural in quantum-information settings where preparation, control, and readout are local rather than global~\cite{Preskill2018quantumcomputingin}. Finite-velocity propagation further motivates a fixed-window formulation of local recovery~\cite{Lieb1972,Nahum2017}.

\begin{figure*}[t]
	\centering
	\includegraphics[width=\textwidth]{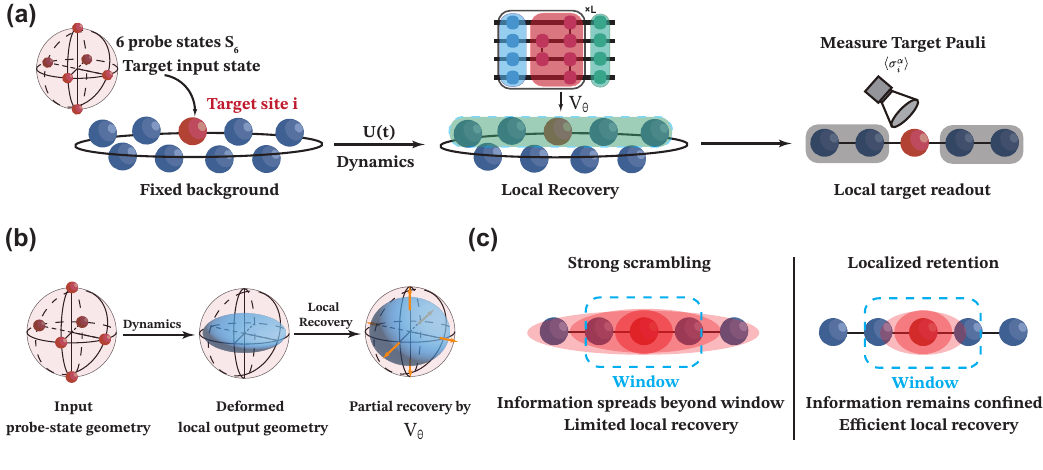}
	\caption{
		Local variational recovery operationalizes finite-window recoverability with strictly local resources.
		(a) Protocol: prepare the target qubit in one of six probe states while the remaining qubits are initialized in a fixed product-state background, evolve under the many-body dynamics of interest, apply a local decoder $V_\theta$ on a radius-$r$ recovery window, and measure target-qubit Pauli operators.
		(b) Channel viewpoint: many-body dynamics deforms the local output geometry of the probe ensemble, and the optimized decoder partially restores it.
		(c) Conceptual distinction: under strong scrambling, relevant information spreads beyond the accessible window and local recovery fails, whereas in a disorder-dominated regime the information remains confined closely enough to permit efficient finite-window recovery.
	}
	\label{fig:protocol}
\end{figure*}

Existing diagnostics answer different questions. Spectral statistics identify crossover windows between ergodic and disorder-dominated behavior~\cite{PhysRevB.75.155111,PhysRevB.82.174411}, but they do not determine where the recoverable information resides. Observable, echo, and scrambling diagnostics probe target-site persistence, reversibility, or operator growth~\cite{Schreiber2015,Smith2016,PhysRevLett.120.050507,PhysRevE.65.036208,PhysRevB.96.014202,Prosen2007,Gaerttner2017,García-Mata:2023}, but they do not optimize a constrained window-to-target decoder. Tomographic channel diagnostics reconstruct local channel content, and broader reconstruction-based workflows enlarge that toolbox~\cite{Gross2010QST,AnWuYangZhouZeng2024}, but they do not directly ask what bounded local control can return to the target qubit. We therefore define recovery with respect to an explicit access model.

We address that gap by formulating finite-window recoverability as a locality-constrained channel-recovery task, in the broader spirit of quantum recovery-map and approximate-correction formulations~\cite{BarnumKnill2002,BenyOreshkov2010,FawziRenner2015,Sutter2016}. The access model is deliberately narrow: local state preparation, many-body evolution, bounded-depth control on a chosen recovery window, and target-qubit Pauli readout. Within that model we introduce local variational recovery (LVR), a shallow local-decoder protocol that estimates recoverability without requiring tomography of the recovery window. LVR therefore implements a bounded local-control ansatz for an operational recovery task. Related local-circuit protocols have also used single-qubit disentangling performance to probe many-body structure, but for a different task than the present channel-recovery problem~\cite{AnCaoXuZhou2024}.

We compare three channel-level quantities. Let $\mathcal{E}_0$ denote the effective channel from the input qubit to the target-site output and $\mathcal{E}_r$ the corresponding channel to the full radius-$r$ recovery window. The optimal target-only score $Q^{\mathrm{opt}}_0$ measures what remains recoverable from the target output alone, the shallow score $Q^{\mathrm{shallow}}_r$ measures what bounded-depth local control actually recovers from the finite window, and the same-window optimum $Q^{\mathrm{opt}}_r$ measures what is recoverable in principle from that same window. The certified gain
\begin{equation}
\Delta^{\mathrm{cert}}_r = Q^{\mathrm{shallow}}_r - Q^{\mathrm{opt}}_0
\label{eq:intro_delta_cert}
\end{equation}
then asks whether recoverable information exists outside the target site but inside the chosen window, while the shallow-accessibility fraction
\begin{equation}
\eta_r = \frac{Q^{\mathrm{shallow}}_r - Q^{\mathrm{opt}}_0}{Q^{\mathrm{opt}}_r - Q^{\mathrm{opt}}_0}
\label{eq:intro_eta}
\end{equation}
asks how much of that same-window recoverable content is already accessible to bounded-depth local control.

As a first benchmark we study a disordered kicked-Ising Floquet chain, where periodic driving, interactions, and disorder compete directly~\cite{PhysRevLett.114.140401,PhysRevLett.115.030402}. The broader Floquet setting is shaped by heating, localization, and prethermalization considerations~\cite{Bukov2015,Abanin2016,Mori2018}. This benchmark provides a useful test bed because weak disorder allows information to escape beyond a small recovery window, whereas stronger disorder can preserve substantial local memory. Beyond the exact-diagonalization benchmark, we also test whether positive certified gain persists when the same radius-2 task is generated by substantially longer chains using an independent tensor-network backend.

Our first result is a channel-level hierarchy for local recovery. For a five-site decoder we find \(Q^{\mathrm{opt}}_0 < Q^{\mathrm{shallow}}_2 < Q^{\mathrm{opt}}_2\), with positive paired-disorder certified gain in the crossover regime and substantial shallow-accessibility fractions on the full exact benchmark. This establishes that loss of on-site recoverability does not imply loss from the nearby window, and that a bounded-depth decoder can access a nontrivial part of the same-window recoverable content.

We then use this hierarchy to design a stricter repair benchmark. In the intermediate-disorder window where target-only recovery has weakened but finite-window recovery remains accessible, we delete and reset the original target carrier and ask whether the surrounding halo can repair the input state. At a frozen held-out radius-3 working point, a depth-8 decoder recovers the deleted carrier with median \(F_{\mathrm{avg}}=0.758\), above the single-qubit classical benchmark \(2/3\), and its paired recovery score exceeds optimal CPTP decoders restricted to every one-, two-, and three-site halo subwindow. This converts the finite-window hierarchy from a diagnostic of off-site recoverable memory into an explicit local repair stress test.

This paper develops a channel-level benchmark for finite-window recoverability under a fixed local-access model. Section II defines the effective channels, recovery benchmarks, and certification quantities. Section III first establishes the exact \(N=12\) recovery hierarchy, then uses the hierarchy-selected intermediate-disorder window to define and test a carrier-deletion repair task, and finally compares the recovery signal with target-site persistence and the tomographic channel-information comparator. Section IV tests whether the positive radius-2 gain persists on longer-chain tensor-network outputs. Methods records the reproducibility-critical numerical ingredients, including the held-out carrier-deletion protocol, and the appendices collect proofs and supporting controls.
\section{Operational Framework and Certification}

\subsection{Access model and effective channels}

Fix a target site $i$ on a lattice $\mathcal{L}$. For an integer recovery radius $r$, let
\begin{equation}
\Lambda_r(i)=\{j\in\mathcal{L}:\operatorname{dist}_{\mathcal{L}}(i,j)\le r\},
\end{equation}
where $\operatorname{dist}_{\mathcal{L}}$ is the lattice distance. For the periodic $N$-site chains used in the exact benchmark this is equivalently $\Lambda_r(i)=\{(i+d)\bmod N:d=-r,\ldots,r\}$, while for open chains we intersect the same interval with the physical lattice. In particular, $\Lambda_0(i)=\{i\}$.
We initialize the target qubit in an input state $|\psi\rangle$ and all remaining degrees of freedom in a fixed product-state background $|\phi_{\mathrm{bg}}\rangle$, so that
\begin{equation}
|\Psi_{\mathrm{in}}(\psi)\rangle = |\psi\rangle_i \otimes |\phi_{\mathrm{bg}}\rangle_{\bar i}.
\end{equation}
Let $U$ denote the many-body evolution used in the benchmark or application of interest. The target-input channel to the target site alone and to the full radius-$r$ window are then
\begin{equation}
\begin{aligned}
\mathcal{E}_0(\rho)
&=
\mathrm{Tr}_{\bar i}\!\left[
U\,
\bigl(\rho_i\otimes |\phi_{\mathrm{bg}}\rangle\langle\phi_{\mathrm{bg}}|_{\bar i}\bigr)
\right.
\\
&\qquad\left.
\times U^\dagger
\right],
\end{aligned}
\end{equation}
\begin{equation}
\begin{aligned}
\mathcal{E}_r(\rho)
&=
\mathrm{Tr}_{\overline{\Lambda_r(i)}}\!\left[
U\,
\bigl(\rho_i\otimes |\phi_{\mathrm{bg}}\rangle\langle\phi_{\mathrm{bg}}|_{\bar i}\bigr)
\right.
\\
&\qquad\left.
\times U^\dagger
\right],
\end{aligned}
\end{equation}
The overlines in $\bar i$ and $\overline{\Lambda_r(i)}$ denote complements inside the full system. A shallow layered local circuit $V_\theta$ of depth $L$ acting only on $\Lambda_r(i)$ induces a window-to-target CPTP map
\begin{equation}
\mathcal{D}_{\theta,r,L}(\omega)
=
\mathrm{Tr}_{\Lambda_r(i)\setminus\{i\}}
\!\left[V_\theta\,\omega\,V_\theta^\dagger\right].
\end{equation}
The radius $r$ is the locality budget and the layer number $L$ is the decoder-expressivity budget. Throughout, we use the phrase ``local quantum information'' operationally: it refers to recoverability of the target-input channel under the specified probe ensemble, background preparation, and local-access model, rather than to a background-independent property of the many-body dynamics.

\subsection{Shallow score and exact local comparators}

To evaluate recovery on the target qubit without full tomography, we probe the six cardinal states
\begin{equation}
\mathcal{S}_6 = \{|0\rangle, |1\rangle, |\pm\rangle, |\pm i\rangle\},
\end{equation}
and for any recovered single-qubit channel $\mathcal{N}$ define
\begin{equation}
F_6(\mathcal{N})
=
\frac{1}{6}\sum_{\psi\in \mathcal{S}_6}
\bra{\psi}\mathcal{N}\!\left(|\psi\rangle\langle\psi|\right)\ket{\psi},
\end{equation}
\begin{equation}
Q(\mathcal{N})=2F_6(\mathcal{N})-1.
\end{equation}
The fixed-decoder LVR score is then
\begin{equation}
Q_{\mathrm{LVR}}(\theta;r,L)
=
Q\!\left(\mathcal{D}_{\theta,r,L}\circ \mathcal{E}_r\right),
\end{equation}
whose equivalent Pauli-expectation representation is given in Methods. The ansatz-limited optimum is
\begin{equation}
Q^{\mathrm{ansatz}}_{r,L}
=
\max_\theta\,Q\!\left(\mathcal{D}_{\theta,r,L}\circ \mathcal{E}_r\right).
\end{equation}
Numerically, the reported shallow score is
\begin{equation}
Q^{\mathrm{shallow}}_{r,L}
=
Q\!\left(\mathcal{D}_{\hat\theta,r,L}\circ \mathcal{E}_r\right),
\end{equation}
where $\hat\theta$ denotes the best parameter set returned by the fixed training protocol specified in Methods. When the reference depth $L=6$ is fixed, we suppress the depth index and write \(Q^{\mathrm{shallow}}_r\). Restart-robustness checks are reported in Appendix \ref{app:controls}, Fig.~\ref{fig:controls_appendix}.
For exact numerical benchmarking we also define two unrestricted CPTP optima:
\begin{equation}
Q^{\mathrm{opt}}_0
=
\max_{\mathcal{D}_0\ \mathrm{CPTP}}
\Big[2F_6(\mathcal{D}_0\circ \mathcal{E}_0)-1\Big],
\end{equation}
\begin{equation}
Q^{\mathrm{opt}}_r
=
\max_{\mathcal{D}_r\ \mathrm{CPTP}}
\Big[2F_6(\mathcal{D}_r\circ \mathcal{E}_r)-1\Big].
\end{equation}
Here $Q^{\mathrm{opt}}_0$ asks how much memory remains recoverable from the target output alone, $Q^{\mathrm{shallow}}_r$ how much can be recovered by locality-constrained shallow control, and $Q^{\mathrm{opt}}_r$ how much is present inside the accessible window irrespective of shallow decodability. Only $Q^{\mathrm{shallow}}_r$ is the proposed operational protocol; $Q^{\mathrm{opt}}_0$ and $Q^{\mathrm{opt}}_r$ are exact numerical comparators under the same locality restriction.

\subsection{Certification logic and derived quantities}

\mainstatement{Identity}{Six-state estimator}{prop:sixstate_avg}
The six cardinal states $\mathcal{S}_6$ form a qubit state 2-design. Hence, for any recovered single-qubit channel $\mathcal{N}$,
\begin{equation}
\begin{aligned}
F_6(\mathcal{N})
&=
\frac{1}{6}\sum_{\psi\in \mathcal{S}_6}
\bra{\psi}\mathcal{N}\!\left(|\psi\rangle\langle\psi|\right)\ket{\psi}
\\
&=
\int d\psi\,
\bra{\psi}\mathcal{N}\!\left(|\psi\rangle\langle\psi|\right)\ket{\psi}
\\
&\equiv
F_{\mathrm{avg}}(\mathcal{N},\mathrm{id}) .
\end{aligned}
\label{eq:sixstate_identity_main}
\end{equation}
We report the rescaled score $Q(\mathcal{N})=2F_{\mathrm{avg}}(\mathcal{N},\mathrm{id})-1$, so that perfect recovery gives $Q=1$ and the completely depolarizing channel gives $Q=0$. The reference value $F_{\mathrm{avg}}=2/3$ is the single-qubit classical benchmark: it is the largest Haar-averaged fidelity achievable for an unknown qubit by a measure-and-prepare, equivalently entanglement-breaking, recovery strategy~\cite{MassarPopescu1995}. Hence a recovered qubit channel with $F_{\mathrm{avg}}>2/3$ cannot be explained as purely classical state estimation followed by re-preparation. In the rescaled variable this benchmark is $Q_{\mathrm{cl}}=2(2/3)-1=1/3$. We use this absolute benchmark only to interpret the reset-erasure repair task; the finite-window certification hierarchy itself is based on the relative witness $\Delta^{\mathrm{cert}}_r=Q^{\mathrm{shallow}}_r-Q^{\mathrm{opt}}_0$.

\mainstatement{Observation}{No finite-window gain from pure CPTP post-processing}{prop:postprocess_no_gain}
If there exists a CPTP map $\mathcal{T}_r$ such that
\begin{equation}
\mathcal{E}_r = \mathcal{T}_r \circ \mathcal{E}_0 ,
\label{eq:postprocess_assumption_main}
\end{equation}
then
\begin{equation}
Q^{\mathrm{opt}}_r = Q^{\mathrm{opt}}_0 .
\label{eq:postprocess_nogain_main}
\end{equation}
Therefore any observed $Q^{\mathrm{shallow}}_r > Q^{\mathrm{opt}}_0$ certifies recoverable information outside the target site but inside the chosen radius-$r$ window (see details in Appendix \ref{app:lvr_details} and Appendix \ref{app:onsite_cptp}).

The operational content of the construction is not exhausted by this binary certification statement. The certified finite-window gain
\begin{equation}
\Delta^{\mathrm{cert}}_r = Q^{\mathrm{shallow}}_r - Q^{\mathrm{opt}}_0
\label{eq:delta_cert_main}
\end{equation}
asks whether the accessible window contains recoverable information beyond the target-site output alone. When the optimal same-window benchmark is also available, we define the shallow-accessibility fraction
\begin{equation}
\eta_r = \frac{Q^{\mathrm{shallow}}_r - Q^{\mathrm{opt}}_0}{Q^{\mathrm{opt}}_r - Q^{\mathrm{opt}}_0},
\label{eq:eta_main}
\end{equation}
which asks how much of that same-window recoverable content is already reached by bounded-depth local control. We therefore interpret $\Delta^{\mathrm{cert}}_r$ as an existence quantity and $\eta_r$ as an accessibility quantity for the same fixed window.

\paragraph*{Convention for the shallow-accessibility fraction.}
We report $\eta_r$ only on instances for which the optimal same-window benchmark $Q^{\mathrm{opt}}_r$ was explicitly evaluated and the denominator $Q^{\mathrm{opt}}_r-Q^{\mathrm{opt}}_0$ is numerically positive. For the exact $N=12$ kicked-Ising benchmark, this now corresponds to the full 100-realization disorder grid. In the larger-$N$ tensor-network study, by contrast, $\eta_2$ is shown only on the subset on which the same-window SDP was explicitly solved. 

\noindent
As a tomographic comparator, we also evaluate a channel-information increment from the reconstructed Choi states. Let $\Omega_{R\Lambda_r}=({\mathrm{id}}_R\otimes\mathcal{E}_r)(\Phi^+_{Ri})$, where $R$ is a reference qubit maximally entangled with the input, and let $\Omega_{RA}$ denote its marginal on an output region $A\subseteq\Lambda_r(i)$. For any such region, define $I_{\mathrm{c}}(R\rangle A)_\Omega=S(\Omega_A)-S(\Omega_{RA})$. We then set
\begin{equation}
\begin{aligned}
\Delta I_{\mathrm{c},r}
&=
I_{\mathrm{c}}(R\rangle \Lambda_r)_\Omega
-
I_{\mathrm{c}}(R\rangle \Lambda_0)_\Omega,
\\
&\qquad
\Lambda_0\equiv \Lambda_0(i)=\{i\}.
\end{aligned}
\end{equation}
This quantity measures the additional coherent channel information in the radius-$r$ output relative to the target-site output. It is an entropic, tomography-based comparator, not an optimized recovery score; the latter is $Q^{\mathrm{opt}}_r$. For the trace-preserving single-qubit input channel considered here, $S(\Omega_R)=1$ bit is independent of the output region, so $\Delta I_{\mathrm{c},r}$ equals the corresponding mutual-information increment $\Delta I_r=I(R:\Lambda_r)-I(R:\Lambda_0)$.

\section{Exact Kicked-Ising Benchmark}

\subsection{Model and benchmark settings}

As a first exact benchmark we use a periodically driven disordered kicked-Ising chain with periodic boundary conditions and site indices $j=0,1,\dots,N-1$. One Floquet period is
\begin{widetext}
\begin{equation}
U_F(W,\mathbf h)
=
\exp\!\Big(-i g\,\delta t \sum_{j=0}^{N-1} X_j\Big)\,
\exp\!\Big(-i J\,\delta t \sum_{j=0}^{N-1} Z_j Z_{j+1}\Big)\,
\exp\!\Big(-i \delta t \sum_{j=0}^{N-1} W h_j Z_j\Big),
\label{eq:kickedising_floquet}
\end{equation}
\end{widetext}
with $j+1$ understood modulo $N$ and $h_j\sim \mathrm{Unif}[-1,1]$. We evolve for $D$ periods, $U(D;W,\mathbf h)=\left[U_F(W,\mathbf h)\right]^D$, and unless otherwise stated fix $J=g=1$, $\delta t=0.2$, target site $i=N/2=6$, a N\'eel background on the non-target sites, recovery radius $r=2$, decoder depth $L=6$, and 100 disorder realizations per disorder point. A conventional adjacent-gap-ratio benchmark is retained only as a spectral consistency check on the crossover window rather than as the article's organizing diagnostic~\cite{PhysRevB.75.155111,PhysRevB.82.174411} (see details in Appendix \ref{app:kicked_ising}).

\subsection{Recoverability hierarchy and paired-disorder certification}

	\begin{figure*}[t]
	    \centering
	    \includegraphics[width=\textwidth]{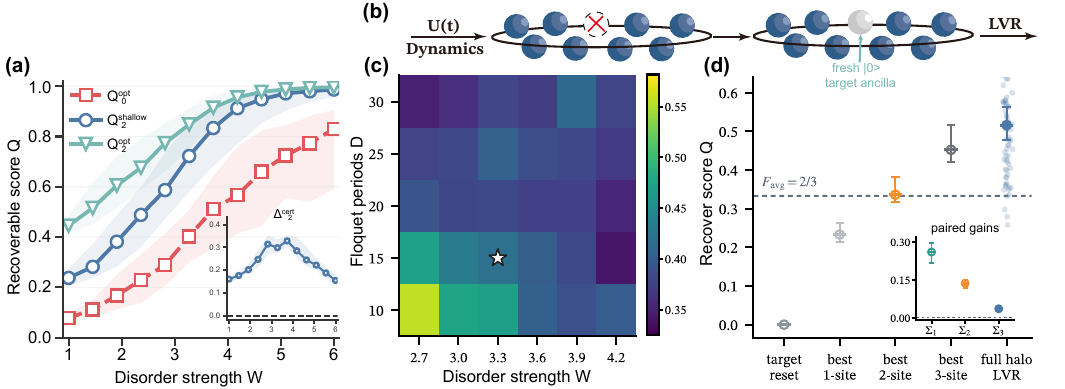}
\caption{
Hierarchy-guided carrier-deletion repair from a finite surrounding halo.
(a) Radius-2 recovery hierarchy in the original task, comparing the target-site optimum $Q^{\mathrm{opt}}_0$, the operational depth-6 score $Q^{\mathrm{shallow}}_2$, and the same-window optimum $Q^{\mathrm{opt}}_2$; the inset shows the certified gain $\Delta^{\mathrm{cert}}_2$.
(b) Carrier-deletion protocol: after the many-body evolution, the target carrier is reset and the radius-3 surrounding halo is decoded back to the target register.
(c) Shallow-only reset-erasure selection scan over disorder strength and Floquet depth; the star marks the frozen held-out working point.
(d) Held-out subwindow test at the frozen point. The full-halo shallow decoder exceeds the single-qubit classical benchmark $Q=1/3$ ($F_{\mathrm{avg}}=2/3$) and outperforms the best one-, two-, and three-site halo-subwindow CPTP counterfactuals; the inset shows the paired gains.
}
	 \label{fig:certified_hierarchy}
	\end{figure*}

Figure~\ref{fig:certified_hierarchy}(a) compares three recovery levels on the same radius-2 channel. The target-site optimum $Q^{\mathrm{opt}}_0$ measures what remains recoverable from the target output alone. The shallow score $Q^{\mathrm{shallow}}_2$ measures what the depth-6 local decoder actually recovers from the same five-site window. The same-window optimum $Q^{\mathrm{opt}}_2$ measures what is recoverable in principle from that unchanged window. The hierarchy
\begin{equation}
Q^{\mathrm{opt}}_0 < Q^{\mathrm{shallow}}_2 < Q^{\mathrm{opt}}_2
\label{eq:hierarchy_main}
\end{equation}
therefore separates off-site recoverable content from its accessibility within the same locality budget.

Across the full exact benchmark, $\Delta^{\mathrm{cert}}_2$ and $\eta_2$ characterize two numerically distinct features of the same radius-2 channel: the excess recovery over the on-site optimum and the fraction of the same-window optimum reached by the depth-6 ansatz. The certified-gain inset in Fig.~\ref{fig:certified_hierarchy}(a) visualizes the positive gain that guides the repair search, while the shallow-accessibility fractions quoted below are computed from the same full exact benchmark. The median certified gain is positive across the full disorder scan and is largest in the crossover regime; pointwise paired Wilcoxon signed-rank tests together with bootstrap intervals for the median gain and the positive fraction support the same paired-disorder conclusion across the full scan. The same five-site shallow decoder already accesses a substantial fraction of the locally available memory throughout the crossover-to-strong-disorder window on the full exact benchmark. The median shallow-accessibility fraction rises from about $0.50$ at $W\approx 2.36$ to about $0.94$ by $W\approx 5.09$, while reaching about $0.83$ already near the certified-gain peak at $W\approx 3.73$.

Thus \(Q^{\mathrm{shallow}}_2>Q^{\mathrm{opt}}_0\) establishes off-site recoverable memory within \(\Lambda_2\), while the residual gap \(Q^{\mathrm{opt}}_2-Q^{\mathrm{shallow}}_2\), summarized by \(\eta_2\), quantifies how much of that same-window advantage is reached by the fixed depth-6 decoder. The certified-gain profile therefore identifies an intermediate-disorder operating window in which the target-site channel alone is no longer the relevant memory carrier, but the nearby window still contains shallow-accessible recoverable content. We use this window in the next subsection to impose a stricter carrier-deletion stress test.

\subsection{Carrier-deletion repair from a surrounding halo}

The finite-window hierarchy suggests a sharper repair question: if the target-site channel has ceased to be the sole recoverable carrier, can the original carrier be removed altogether and repaired from the surrounding halo? We test this by replacing the target qubit in a radius-$r$ output window, with $r=3$ in the held-out benchmark below, by a fresh state,
\begin{equation}
	\mathcal{R}_i^{(r)}(\omega_{\Lambda_r})=
	|0\rangle\langle0|_i\otimes \mathrm{Tr}_i\omega_{\Lambda_r}.
\end{equation}
After this reset-erasure map, we call $H_r(i)=\Lambda_r(i)\setminus\{i\}$ the radius-$r$ surrounding halo. The decoder acts on the reset window $H_r(i)\cup\{i\}$, but all input-dependent information available for repair resides in $H_r(i)$. Any successful recovery must therefore read a remnant stored in the surrounding halo and write it back into the reset target.
This carrier-deletion protocol is schematized in Fig.~\ref{fig:certified_hierarchy}(b).

We define the carrier-deletion score of a shallow decoder by
\begin{equation}
Q^{\mathrm{del}}_{r,L}
=
Q\!\left(\mathcal{D}_{\hat\theta^{\mathrm{del}},r,L}\circ \mathcal{R}_i^{(r)}\circ \mathcal{E}_r\right),
\end{equation}
where $\hat\theta^{\mathrm{del}}$ is the best parameter set returned by optimizing the reset-erasure recovery objective for the specified $r$ and $L$.
To test whether the recovered channel can be explained by a smaller displaced subregion, we also define
\begin{equation}
\begin{aligned}
Q^{\mathrm{opt}}_{m\mathrm{-sub}}
&=
\max_{\substack{A\subset \Lambda_r(i)\setminus\{i\}\\ |A|=m}}
\max_{\mathcal{D}_A\ \mathrm{CPTP}}
\\
&\quad
Q\!\left(
\mathcal{D}_A
\circ \mathrm{Tr}_{\Lambda_r(i)\setminus A}
\circ \mathcal{R}_i^{(r)}
\circ \mathcal{E}_r
\right),
\end{aligned}
\end{equation}
and the paired subwindow-exclusion gain
\begin{equation}
\Sigma_m
=
Q^{\mathrm{del}}_{r,L}
-
Q^{\mathrm{opt}}_{m\mathrm{-sub}}.
\end{equation}
These subwindow optima are not the full-halo CPTP ceiling; they are restricted counterfactuals asking whether some small halo subregion could already account for the recovered channel.

Figure~\ref{fig:certified_hierarchy}(c) shows the shallow-only selection scan on a disjoint selection set. In the intermediate-disorder region identified by the hierarchy, the reset-erasure score remains above the single-qubit classical-benchmark line $Q=1/3$ over an extended range of $W$ and Floquet depth $D$. We therefore freeze the representative point $W=3.3$, $D=15$, $r=3$, and decoder depth $L=8$, and use it only for held-out testing.

Figure~\ref{fig:certified_hierarchy}(d) reports the held-out subwindow test at the frozen point. On the 80 held-out disorder realizations, the full-halo shallow decoder reaches median \(Q=0.516\), corresponding to \(F_{\mathrm{avg}}=0.758\). The 95\% bootstrap interval \(Q\in(0.478,0.567)\) lies above the single-qubit classical benchmark \(Q=1/3\), equivalently \(F_{\mathrm{avg}}=2/3\), and 95\% of held-out realizations exceed this benchmark. Thus, even after complete deletion of the target carrier, the surrounding halo retains an input-dependent channel remnant that a shallow local decoder can write back into the reset target register.

The same held-out comparison asks whether this recovery is only a displaced-carrier effect. For each held-out realization, every one-, two-, and three-site halo subwindow is given its most favorable possible decoder, an unrestricted CPTP recovery map. These are deliberately strong counterfactuals: they ask whether some small part of the halo could by itself account for the recovered channel. The best one-, two-, and three-site subwindow CPTP medians are \(Q=0.233\), \(0.336\), and \(0.452\), respectively, while the full-halo shallow decoder gives \(Q=0.516\). In paired statistics, the corresponding gains are \(\Sigma_1=0.260\), \(\Sigma_2=0.136\), and \(\Sigma_3=0.035\), with all 95\% bootstrap intervals strictly positive. The three-site separation is modest, but it is statistically clean. The recovery is therefore not explained by the target register, by a single displaced neighbor, or by any tested small halo subwindow, even when those subwindows are granted unrestricted CPTP recovery maps.

The result shows that the LVR hierarchy can identify a finite surrounding halo from which a deleted carrier can be actively repaired under explicit locality and depth constraints.

\subsection{Distinction from target-site persistence and tomographic channel information}

To compare this hierarchy with more familiar diagnostics, we next examine target-site persistence and the tomographic channel-information comparator under the same disorder scan.

\begin{figure*}[t!]
	    \centering
	    \includegraphics[width=\textwidth]{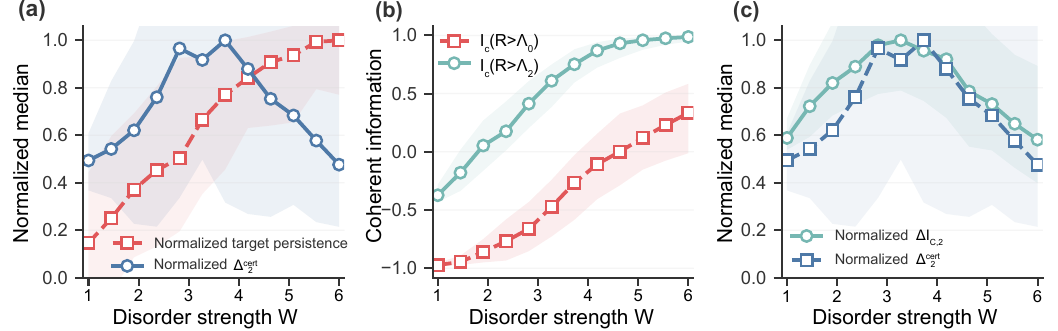}
	    \caption{
	        Alternative diagnostics do not isolate the same operational distinction.
	        (a) Disorder profiles of target-site persistence and certified finite-window gain after normalizing each median curve by its own maximum, making the shape difference directly visible on a common scale.
	        (b) Reconstructed-channel coherent information for the target site and the five-site window.
	        (c) Normalized coherent-information increment $\Delta I_{\mathrm{c},2}$ compared with the normalized certified gain.
	        The local observable and the tomographic comparator both respond to memory retained near the target, but neither isolates the locality-constrained channel-recovery question addressed by $Q^{\mathrm{shallow}}_2$ and $\Delta^{\mathrm{cert}}_2$.
	    }
	    \label{fig:alternative_diagnostics}
	\end{figure*}

Fig.~\ref{fig:alternative_diagnostics} (a) shows why the signal is not reducible to a standard local observable. The target-site magnetization retention $\langle Z_i(D)\rangle_{|0\rangle}$ grows almost monotonically with disorder and therefore tracks direct target persistence, much like the local-imbalance observables commonly used in localization experiments~\cite{Schreiber2015,Smith2016,PhysRevLett.120.050507}. By contrast, the certified finite-window gain peaks in the intermediate regime where on-site recovery has already weakened but finite-window recovery remains strong.

 Fig.~\ref{fig:alternative_diagnostics} (b) and (c) address the tomographic comparator directly. The reconstructed-channel coherent-information increment $\Delta I_{\mathrm{c},2}$ counts channel content retained anywhere inside the accessible window, whereas $\Delta^{\mathrm{cert}}_2$ asks what actually beats the optimal on-site baseline under bounded local decoding. The two quantities can therefore track overlapping disorder windows while answering different access questions.

Appendix~\ref{app:channel_info} gives this distinction a more microscopic form through the operator shell weight (see Fig.~\ref{fig:operator_shell_weights}). Local support retained inside $\Lambda_2$ is necessary but not sufficient for shallow target recovery: the shell weight can continue increasing even after $\Delta^{\mathrm{cert}}_2$ has turned over.

\section{Longer-Chain Embedding of the Radius-2 Recovery Witness}

This section addresses a narrower but concrete question: whether the positive radius-2 certified gain is an artifact of the $N=12$ exact-diagonalization setting in which it is first established. To test that point, we keep the recovery task unchanged and instead embed it in substantially longer chains, open boundaries, and an independent tensor-network backend.

What changes here is the global many-body generator of the local channel, not the size of the recovery problem. The tensor-network backend supplies reduced outputs on the same radius-2 window, after which $Q^{\mathrm{shallow}}_2$, $Q^{\mathrm{opt}}_0$, and, on the subset where the same-window SDP is explicitly solved, $Q^{\mathrm{opt}}_2$ are evaluated exactly as in the exact benchmark. The point of the section is therefore persistence of positive radius-2 gain under longer-chain embedding, not an algorithmic scalability claim for window sizes that grow with system size.

\subsection{Tensor-network backend and subset protocol}

For each disorder realization, disorder strength, and probe input, we evolve the kicked-Ising chain with an open-boundary MPS/TEBD backend and then trace the resulting many-body state down to the recovery window, following the same general strategy that has enabled tensor-network studies of disorder-dominated dynamics beyond exact diagonalization~\cite{Doggen2021,PhysRevB.107.115132}. In the verified implementation, one Floquet period is applied directly as an $RX$ layer, a nearest-neighbor $ZZ$ layer, and onsite $RZ$ layers, so the tensor-network approximation enters through bond truncation rather than an additional Trotter breakup of the Floquet step. All recovery quantities are subsequently computed from those reduced outputs alone.

\subsection{Large-N witness behavior}

\begin{figure*}[htbp]
    \centering
    \includegraphics[width=0.95\textwidth]{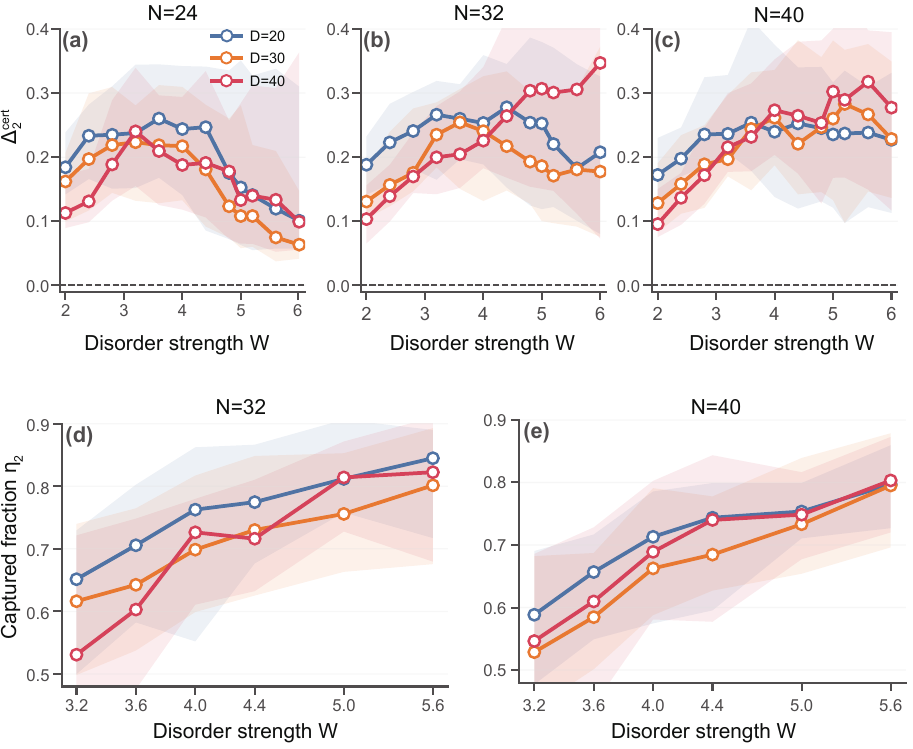}
    \caption{
        Larger-system tensor-network deployment of the finite-window recoverability framework.
        (a)--(c) Median certified finite-window gain $\Delta^{\mathrm{cert}}_2=Q^{\mathrm{shallow}}_2-Q^{\mathrm{opt}}_0$ versus disorder strength for $N=24,32,40$ at representative Floquet depths $D=20,30,40$; shaded bands denote the IQR over 32 disorder realizations.
    }
    \label{fig:largeN_extension}
\end{figure*}

Figure~\ref{fig:largeN_extension} addresses the deployment question directly. For $N=24$, the strongest intermediate-disorder signal occurs near $W\approx 3.2$--$3.6$ at $D=20$--$40$; for $N=32$, the median gain peaks near $W\approx 4.4$ at $D=20$ and near $W\approx 3.6$ at $D=30$; and for $N=40$, the peak remains in the crossover-to-strong-disorder window. On the subset where the same-window optimum is evaluated, the shallow decoder still accesses a substantial fraction of the locally available memory, with typical $\eta_2$ values of about $0.65$--$0.85$ for $N=32$ and about $0.55$--$0.80$ for $N=40$. These data indicate that the positive radius-2 certified gain is not specific to the $N=12$ exact-diagonalization instance.

\section*{Methods}\phantomsection\label{sec:methods}

\paragraph{Scope of the operational claim.}
The experimentally operational quantity in this work is the shallow finite-window recovery score \(Q^{\mathrm{shallow}}_r\), evaluated from local state preparation, many-body evolution, a depth-\(L\) decoder acting only on \(\Lambda_r(i)\), and six target-qubit Pauli settings. The exact benchmarks \(Q^{\mathrm{opt}}_0\), \(Q^{\mathrm{opt}}_r\), and \(\Delta I_{\mathrm{c},r}\) require reconstructed reduced states and semidefinite or entropic post-processing, and are used as numerical comparators on tractable instances. In the present work, the recovery radius is part of the task definition, and the shallow decoder is the concrete bounded local-control ansatz used to implement that task.

Unless otherwise stated, all exact $N=12$ scans use the same ordered set of 100 disorder realizations at each of the 12 disorder points. The optimal same-window benchmark $Q^{\mathrm{opt}}_2$, and hence the shallow-accessibility fraction $\eta_2$, is now evaluated across that full exact 100-realization data set. In the longer-chain tensor-network deployment, the shallow scans use the full $(W,D)$ grids, whereas $Q^{\mathrm{opt}}_2$ is evaluated only on the subset reported in Fig.~\ref{fig:largeN_extension}. In the verified preset scripts, that subset corresponds to $W=\{3.2,3.6,4.0,4.4,5.0,5.6\}$ and $D=\{10,20,30,40\}$. These subset evaluations sample the same-window benchmark and do not define the shallow-recovery scans themselves.

\paragraph{Model and Floquet evolution.}
The main interacting study uses a periodic chain of $N=12$ spins with target site $i=N/2=6$. Disorder is sampled as independent base fields $h_j\sim \mathrm{Unif}[-1,1]$, rescaled by the plotted disorder strength $W$. One Floquet period is the sequence of uniform $X$ rotations, nearest-neighbor Ising $ZZ$ couplings, and disordered $Z$ rotations given in Eq.~\eqref{eq:kickedising_floquet}, with fixed couplings $J=g=1$ and time step $\delta t=0.2$. Unless otherwise stated, the system is evolved for $D=30$ periods, the non-target qubits are initialized in a N\'eel product state, and disorder is scanned on the 12-point grid $W\in[1,6]$ used throughout the numerical study.

\paragraph{Longer-chain tensor-network backend.}
For the $N=24,32,40$ deployment, the same Floquet sequence is implemented with an open-boundary MPS backend. In the verified scripts, one period is applied directly as an $RX$ layer, a nearest-neighbor $ZZ$ layer, and onsite $RZ$ layers, so the approximation enters through bond truncation. The standard runs use maximum bond dimension $\chi=128$ and singular-value cutoff $10^{-10}$; the saved output also records the realized maximum bond dimension and the maximum discarded weight per period for each instance.

\paragraph{Recovery-window geometry and decoder ansatz.}
For radius $r$, the accessible window is $\Lambda_r(i)=\{i-r,\dots,i+r\}$ on the ring, with site labels understood modulo $N$. The reference benchmark uses $r=2$, so the decoder acts on the five-site window $\Lambda_2(6)=\{4,5,6,7,8\}$ with nearest-neighbor edges inside that window. The decoder is a layered nearest-neighbor circuit of on-site Euler rotations and controlled-$Z$ gates acting only on $\Lambda_r(i)$; its explicit form is given in Appendix~\ref{app:lvr_details}. The main scan uses $L=6$ trainable entangling layers.

\paragraph{Carrier-deletion held-out protocol.}
For the radius-3 carrier-deletion task, the target site is traced out from the radius-3 output window and reinserted as a fresh \(|0\rangle\) ancilla, i.e., by applying $\mathcal{R}_i^{(3)}$ before the shallow decoder. The operating window was selected from a shallow-only scan on realization IDs 0--19. The quoted held-out result uses the frozen point \(W=3.3\), \(D=15\), \(r=3\), and \(L=8\) on realization IDs 20--99. The subwindow counterfactuals optimize over all one-, two-, and three-site halo subsets using unrestricted CPTP maps; these are restricted subwindow benchmarks and not the full-halo CPTP optimum.

\paragraph{Six-state estimator and Pauli-expectation implementation.}
The probe set is $\mathcal{S}_6=\{|0\rangle,|1\rangle,|\pm\rangle,|\pm i\rangle\}$. The operational score is evaluated from six target-qubit Pauli expectations, and by Identity~\ref{prop:sixstate_avg} it equals $2F_{\mathrm{avg}}(\mathcal{N},\mathrm{id})-1$ for the recovered single-qubit channel $\mathcal{N}$. The explicit estimator and its derivation are given in Appendix~\ref{app:lvr_details}.

\paragraph{Exact CPTP baseline and same-window SDP optimization.}
The exact numerical baselines use the same six probe outputs but maximize over all CPTP decoders compatible with the available output system, using a standard Choi-matrix semidefinite-program formulation of optimal recovery~\cite{FletcherShorWin2007}. The explicit optimization definitions and proof details are given in Appendix~\ref{app:onsite_cptp}. In the verified computations, the on-site SDPs use SCS with $\epsilon=10^{-6}$ and \texttt{max\_iters}=10000, while the optimal same-window benchmark uses SCS with $\epsilon=10^{-5}$ and \texttt{max\_iters}=20000 across the full exact 100-realization benchmark. The positive-gain statements in the main text concern median effects of order $10^{-1}$, well above these nominal SDP tolerances.

\paragraph{Tomographic channel-information benchmark.}
The same six probe outputs reconstruct the target-input channel to the accessible window, from which we evaluate $I(R:\Lambda_r)$, $I_{\mathrm{c}}(R\rangle \Lambda_r)$, and the site-to-window increment $\Delta I_{\mathrm{c},r}$. The reconstruction formulas are given in Appendix~\ref{app:channel_info}. For the single-qubit input used here, $\Delta I_{\mathrm{c},r}$ coincides numerically with the mutual-information increment, but it remains an exact tomographic comparator rather than the optimal same-window benchmark for recovery.

\paragraph{Paired-disorder statistics, optimizer settings, and training budget.}
The primary radius-$2$ shallow-decoder benchmark and the on-site baseline are aligned on the same 100 disorder realizations and the same 12-point disorder grid. The plotted radius-$2$ score is the implemented decoder value $Q^{\mathrm{shallow}}_2=Q(\mathcal{D}_{\hat\theta,2,6}\circ\mathcal{E}_2)=1-\mathcal{L}^\star_{r=2}$, where \(\hat\theta\) is the best parameter set returned by the fixed training protocol below. Paired-realization statistics use one-sided Wilcoxon signed-rank tests (\texttt{alternative="greater"}, \texttt{zero\_method="pratt"}) and percentile bootstrap intervals from $10^4$ paired resamples for both the median certified gain and the positive fraction. The decoder optimization uses Adam with learning rate $0.03$, 300 optimization steps, and Gaussian angle initialization of width $0.1$ for each disorder realization. The layer-scan control uses the same training budget for $L=2,4,6,8,10$, and the matched depth scan uses depths $D=10,20,30,40,60$ on 20 paired disorder realizations together with CLARABEL-based CPTP subproblems. Appendix \ref{app:controls}, Fig.~\ref{fig:controls_appendix}(b), reports the corresponding multi-start robustness check.

\paragraph{Finite-shot and gate-miscalibration models.}
The finite-shot retraining control uses the same six target-qubit settings as the ideal objective. In the training model, a symmetric readout error $p_{\mathrm{ro}}$ damps each ideal expectation value by the factor $1-2p_{\mathrm{ro}}$, and the finite-shot estimate is modeled by adding Gaussian noise with variance $(1-\langle O\rangle^2)/M$ for $M$ shots in that Pauli setting, clipped to the physical interval $[-1,1]$. The scan uses shot budgets $M=256,1024,4096$, readout-error values $p_{\mathrm{ro}}=0,0.02$, 100 disorder realizations, and the same 300-step Adam budget as the ideal benchmark. The coherent deployment-miscalibration control perturbs the trained radius-$2$ decoder on the reduced five-site output states: every single-qubit Euler angle is rescaled by $1+\delta_{1\mathrm q}$, and every entangling gate uses a controlled phase $\pi(1+\delta_{\mathrm{CZ}})$. The sweep uses 12 disorder realizations, $L=6$, 60 optimization steps for the ideal training stage, and four representative calibration settings: an unperturbed decoder, a $2\%$ one-qubit over-rotation, a combined $2\%$ one-qubit plus $2\%\pi$ controlled-phase perturbation, and a combined $5\%$ one-qubit plus $5\%\pi$ controlled-phase perturbation.

\paragraph{Data provenance.}
The exact reduced-window states that underlie $Q^{\mathrm{opt}}_0$, $Q^{\mathrm{opt}}_2$, and $\Delta I_{\mathrm{c},2}$ are stored as six probe-state outputs on the five-site window $(4,5,6,7,8)$, with tensor dimensions corresponding to 100 disorder realizations, 12 disorder points, and six probe inputs. The accompanying data set also contains the derived benchmark arrays for the hierarchy, the tomographic comparator, the finite-shot and gate-miscalibration controls, the layer scan, and the matched depth scan, together with the analysis scripts used to produce the reported figures.

\section{Discussion}

We have formulated local quantum memory as a finite-window channel-recovery benchmark that separates on-site recoverability, same-window recoverability, and the part accessible to bounded-depth local control. Identity~\ref{prop:sixstate_avg} places the six-state score on standard channel-fidelity footing, and Observation~\ref{prop:postprocess_no_gain} gives the channel-level certification logic behind $\Delta^{\mathrm{cert}}_r>0$. In the exact kicked-Ising benchmark, the three quantities $Q^{\mathrm{opt}}_0$, $Q^{\mathrm{shallow}}_r$, and $Q^{\mathrm{opt}}_r$ distinguish what is recoverable from the target site alone, what is recovered by the chosen local decoder, and what is recoverable in principle from the same window. The two derived quantities $\Delta^{\mathrm{cert}}_r$ and $\eta_r$ then distinguish excess window recovery from the fraction of that excess reached by the chosen decoder.

The carrier-deletion benchmark gives the hierarchy a stronger operational interpretation. The target register is reset by construction, so target-only recovery is reduced to the no-information baseline. Nevertheless, the radius-3 surrounding halo supports repair above the single-qubit classical benchmark by a depth-8 local decoder, and the paired subwindow tests rule out explanations based on any one-, two-, or three-site halo subset. This shows that finite-window recovery can be used not only to certify off-site memory, but also to construct a local repair stress test for a deleted carrier.

For the radius-2, depth-6 benchmark studied here, the data show that off-site recoverable memory can remain present after on-site recovery has weakened, and that same-window recoverable content need not be equally accessible to the shallow decoder. Target-site persistence and $\Delta I_{\mathrm{c},2}$ can track overlapping disorder windows while answering different access questions, and the shell-weight analysis shows that local support inside $\Lambda_2$ is necessary but not sufficient for shallow recovery. In the longer-chain tensor-network deployment, the positive radius-2 certified gain persists, and on the subset where the same-window optimum is evaluated the shallow decoder captures a substantial fraction of the available window advantage.

The present analysis is deliberately limited. It studies a single target input qubit, a fixed product-state background, and a layered CZ-plus-Euler decoder ansatz, with $(r,L,D)=(2,6,30)$ as the reference benchmark. The carrier-deletion result remains a finite-size, single-qubit, fixed-background benchmark at a frozen working point; it does not establish a scalable error-correcting code. In the longer-chain data, $Q^{\mathrm{opt}}_2$ is not evaluated on the full $(W,D)$ grid, so the larger-system claim is restricted to persistence of positive radius-2 gain together with subset-level accessibility fractions.

The same construction extends blockwise to multi-qubit inputs: for a $k$-qubit input block $B$, one may define channels from the encoded block to the target block and to the recovery window, replace the six-state probe ensemble by a tomographically complete block ensemble or a unitary-design surrogate, and define $Q^{\mathrm{opt}}_{B,0}$, $Q^{\mathrm{shallow}}_{B,r}$, $Q^{\mathrm{opt}}_{B,r}$, $\Delta^{\mathrm{cert}}_{B,r}$, and $\eta_{B,r}$ analogously. Within that block-general framework, recovery radius, decoder depth, and evolution time become joint control resources whose tradeoffs can be mapped explicitly. This extension elevates finite-window recovery from a single-qubit witness to a general local-decoder benchmark for when off-site quantum memory is not only present in a finite neighborhood, but actively recoverable with specified locality, depth, and time resources.

\section{Data and Code Availability}

The reduced-window outputs, derived benchmark arrays, and analysis scripts used to generate the figures will be made available upon reasonable request. The data include the exact reduced-window states, hierarchy arrays, tomographic comparator, finite-shot and gate-miscalibration controls, layer scans, and matched-depth scans.

\section{ACKNOWLEDGMENTS}
This work is supported by Science, Technology and Innovation Bureau of Shenzhen Municipality.

\appendix
\setcounter{proposition}{0}
\numberwithin{equation}{section}
\makeatletter
\@addtoreset{figure}{section}
\@addtoreset{table}{section}
\@addtoreset{proposition}{section}
\makeatother
\renewcommand{\thefigure}{\thesection\arabic{figure}}
\renewcommand{\thetable}{\thesection\arabic{table}}
\renewcommand{\theproposition}{\thesection\arabic{proposition}}
\renewcommand{\theHsection}{appendix.\Alph{section}}
\renewcommand{\theHfigure}{appendix.\Alph{section}.\arabic{figure}}
\renewcommand{\theHtable}{appendix.\Alph{section}.\arabic{table}}
\renewcommand{\theHequation}{appendix.\Alph{section}.\arabic{equation}}
\renewcommand{\theHproposition}{appendix.\Alph{section}.\arabic{proposition}}

\section{Physical Picture of the Kicked-Ising Benchmark}
\label{app:kicked_ising}

The kicked-Ising benchmark is a periodically driven interacting spin chain in which each term of the Floquet unitary has a distinct physical role. The transverse-field kick $\exp[-ig\delta t\sum_j X_j]$ rotates every spin and promotes mixing between computational-basis configurations; the interaction step $\exp[-iJ\delta t\sum_j Z_jZ_{j+1}]$ generates many-body correlations; and the disordered longitudinal phase $\exp[-i\delta t\sum_j \tilde h_j Z_j]$ frustrates resonant transport by imprinting site-dependent phases. Their competition makes the model a minimal Floquet setting in which heating and disorder confront each other directly~\cite{PhysRevLett.114.140401,PhysRevLett.115.030402,Abanin2016}. From a broader dynamical viewpoint, the present model also sits at the intersection of Floquet heating, operator complexity, and driven-Ising chaos diagnostics~\cite{Bukov2015,Prosen2000,Prosen2007}. In finite chains such as the ones studied here, we therefore interpret the disorder dependence as a crossover from a thermalizing Floquet regime to a localization-dominated regime, rather than as a sharp thermodynamic transition~\cite{PhysRevB.107.115132,PhysRevE.102.062144,PhysRevB.95.155129}. Near that crossover, slow-dynamics ideas developed for nearly localized systems are often more robust than sharp-transition language~\cite{Gopalakrishnan2020}.

The quasienergy adjacent-gap ratio provides a standard spectral benchmark for that crossover~\cite{PhysRevB.75.155111,PhysRevB.82.174411,RevModPhys.91.021001}. Given sorted eigenphases $\phi_\alpha$ of the one-period Floquet unitary, one forms gaps $\delta_\alpha=\phi_{\alpha+1}-\phi_\alpha$ together with the wrap-around gap, and then averages
\begin{equation}
r^{\mathrm{gap}}_\alpha=\frac{\min(\delta_\alpha,\delta_{\alpha+1})}{\max(\delta_\alpha,\delta_{\alpha+1})}.
\end{equation}
For thermalizing Floquet dynamics with time-reversal structure, the level statistics approach the circular-orthogonal-ensemble value $\langle r^{\mathrm{gap}}\rangle_{\mathrm{COE}}\approx 0.53$, whereas Poisson-like quasienergy statistics give $\langle r^{\mathrm{gap}}\rangle_{\mathrm{P}}\approx 2\ln 2-1 \approx 0.386$~\cite{PhysRevB.75.155111,PhysRevB.82.174411}. Our kicked-Ising scan uses the same disorder samples for both the gap-ratio benchmark and the LVR optimization, so the comparison isolates the distinction between a global spectral benchmark and an operational local-recovery protocol rather than differences in disorder averaging.

From the LVR viewpoint, the same crossover can be phrased dynamically. In the weak-disorder regime, initially local information spreads beyond any small decoder window and the optimized loss remains high. In the strong-disorder regime, the drive admits quasi-local memory structures reminiscent of the local-integral-of-motion phenomenology of MBL, so local information remains reconstructible within a bounded neighborhood~\cite{PhysRevLett.111.127201,Huse2014,ImbrieRosScardicchio2017}. Related structural viewpoints based on area-law eigenstates and Fock-space localization are consistent with the existence of robust local memory, but they do not by themselves determine finite-window recoverability under constrained control~\cite{Bauer2013,BuijsmanGritsevCheianov2018}. We therefore retain the gap-ratio comparison only as a finite-size consistency check on where the crossover window lies. The additional information extracted by LVR is not a different $W^\ast(N)$, but the spatial hierarchy between on-site, finite-window, and same-window-optimal recovery, which a spectral statistic does not encode.

To separate disorder-induced confinement from genuinely interacting redistribution, we compare the kicked-Ising scan against a disordered free-fermion Floquet reference in which the Ising interaction is replaced by nearest-neighbor $XX+YY$ hopping while the decoder protocol, disorder distribution, and window geometry are left unchanged. After a Jordan-Wigner mapping, this reference describes single-particle hopping in a disordered potential rather than many-body Ising interactions. The resulting finite-window gain remains nonzero, so the presence of locally recoverable off-site information is not itself unique to the interacting model. What changes is the disorder profile: the interacting scan develops a pronounced intermediate-regime maximum, whereas the free-fermion gain remains broader and is displaced toward larger disorder. The comparison therefore functions as a mechanism control rather than as evidence that finite-window gain is many-body-specific.

\section{Mechanism Control: Free-Fermion Reference}
\label{app:freefermion_control}

\begin{figure*}[t]
    \centering
    \includegraphics[width=0.94\textwidth]{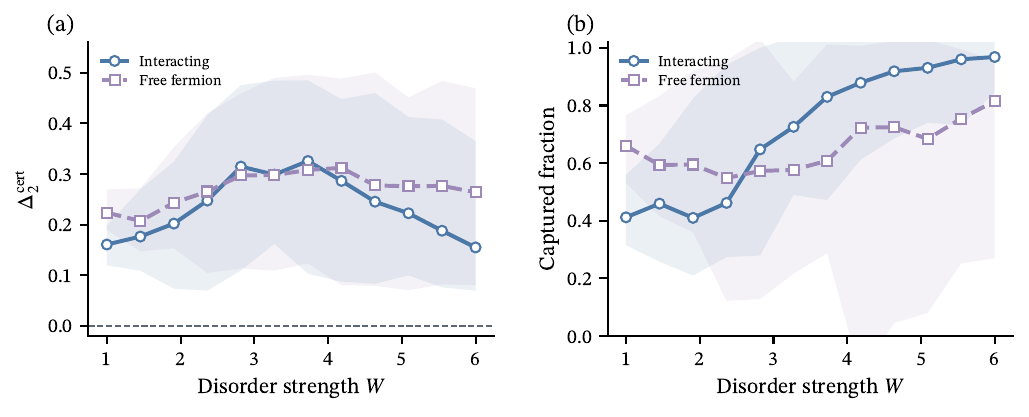}
    \caption{
        Matched free-fermion mechanism control.
        (a) Certified gain $\Delta^{\mathrm{cert}}_2$ in the kicked-Ising chain and in a disordered free-fermion Floquet reference with the same disorder distribution, recovery window, and operational protocol.
        (b) Shallow-accessibility fraction $\eta_2=(Q^{\mathrm{shallow}}_2-Q^{\mathrm{opt}}_0)/(Q^{\mathrm{opt}}_2-Q^{\mathrm{opt}}_0)$ on the full exact interacting benchmark and on the explicitly benchmarked free-fermion same-window sample.
        Solid markers denote disorder medians and shaded bands denote the IQR.
    }
    \label{fig:model_ceiling_control}
\end{figure*}

Figure~\ref{fig:model_ceiling_control} separates the existence of finite-window recoverability from the form it takes in the interacting model. The free-fermion reference also shows positive certified finite-window gain, so the presence of recoverable off-site memory is not uniquely many-body. What the interaction changes is the disorder profile and the efficiency with which shallow local control accesses the memory already present inside the window. The median certified gain peaks near $W\!\approx\!3.73$ in the interacting chain and near $W\!\approx\!4.18$ in the free-fermion reference. Near $W\!\approx\!4.2$, for example, the interacting median shallow-accessibility fraction is about $0.88$ whereas the free-fermion value is about $0.72$, with visibly broader disorder-to-disorder spread in the latter.

\section{Gap-Ratio Consistency Check}
\label{app:wstar_compare}

\begin{figure*}[t]
    \centering
    \includegraphics[width=0.88\textwidth]{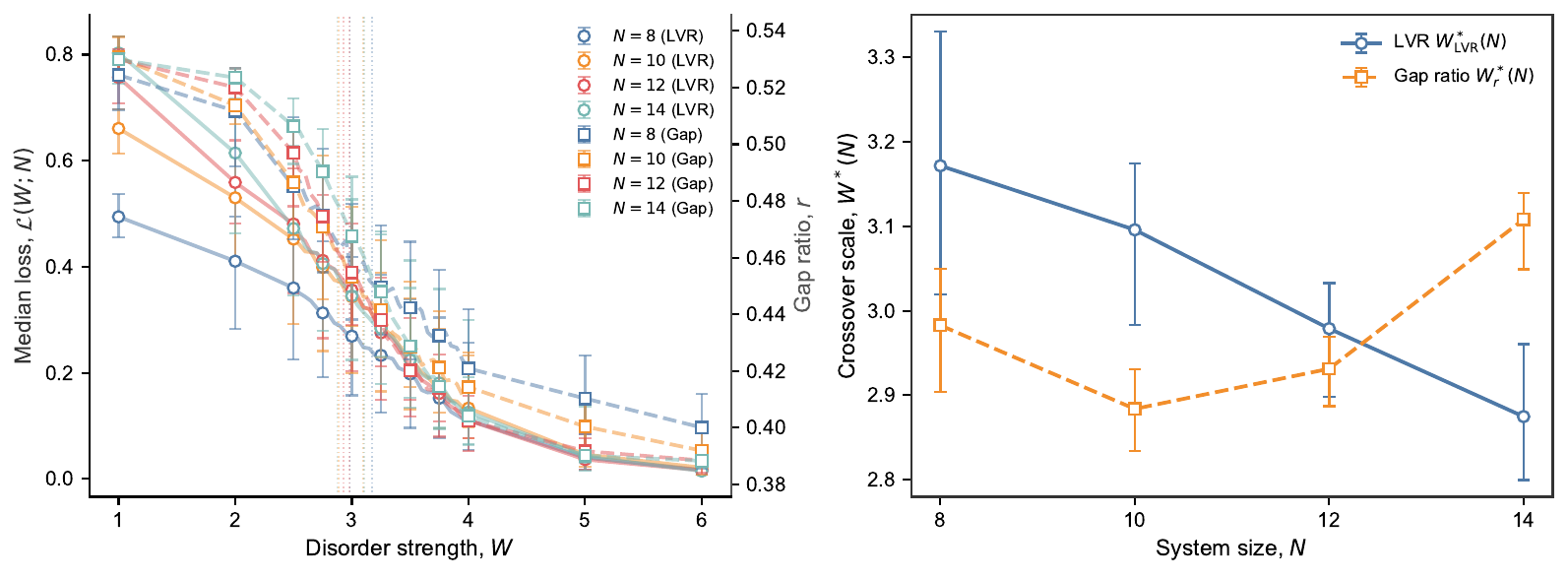}
    \caption{
        Centroid-of-slope extraction of the kicked-Ising crossover scale.
        Both the LVR and adjacent-gap-ratio disorder traces are fit to tanh crossover forms, and the characteristic disorder scale is defined by the centroid of the fitted slope magnitude.
        Within the shared sizes $N=8,10,12,14$, the two diagnostics identify the same finite-size crossover window.
    }
    \label{fig:wstar_v1}
\end{figure*}

As a consistency check, we compare the operational LVR crossover window with the conventional gap-ratio benchmark. The purpose of this section is to show that both observables respond in the same finite-size disorder range; we do not interpret the agreement of $W^\ast(N)$ as the main result, because a single crossover scale is blind to the recoverability profile inside that window. The $W^\ast(N)$ values quoted below are obtained from the same extraction pipeline used to generate Fig.~\ref{fig:wstar_v1}: for each system size and for each benchmark, we fit the disorder trace to a smooth tanh crossover $f_N(W)$ and define
\begin{equation}
W^\ast(N)=
\frac{\int dW\, W\, |f_N'(W)|}
{\int dW\, |f_N'(W)|}.
\end{equation}
In this construction, the tanh fit is used as a smooth finite-size summary of the crossover window; we do not interpret $W^\ast(N)$ as a sharp critical estimator.

\begin{figure*}[t]
    \centering
    \includegraphics[width=0.88\textwidth]{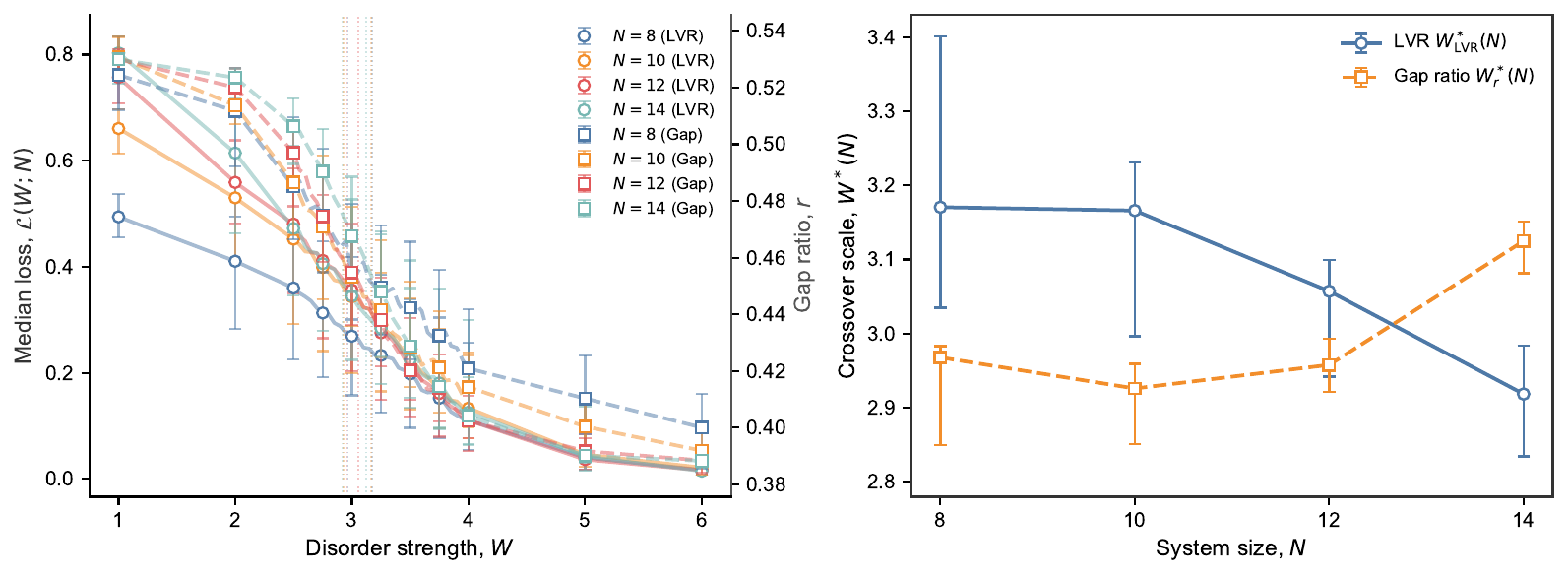}
    \caption{
        Alternative half-height extraction of the kicked-Ising crossover scale.
        Here $W^\ast(N)$ is defined by the linearly interpolated crossing of the midpoint between the low-$W$ and high-$W$ plateaus for both the LVR loss and the gap ratio.
        Within the shared sizes $N=8,10,12,14$, the absolute values shift modestly relative to the centroid-based extraction, but the qualitative agreement between $W^\ast_{\mathrm{LVR}}(N)$ and $W^\ast_{\mathrm{gap}}(N)$ remains intact.
    }
    \label{fig:wstar_v2}
\end{figure*}

Figure~\ref{fig:wstar_v2} shows, for comparison, an alternative half-height extraction based on the linearly interpolated crossing of the midpoint between the low-$W$ and high-$W$ plateaus. The absolute values shift modestly, especially for the decoder curves at intermediate sizes, but the qualitative conclusion is unchanged: the LVR-derived crossover scale and the gap-ratio crossover scale display the same finite-size trend within uncertainty. The smoother centroid-based extraction is used as the default summary, while the half-height construction provides a robustness check.

\section{Effective Recovery-Length Proxy from Radius Scans}
\label{app:rstar_thresholds}

\begin{figure*}[t]
    \centering
    \includegraphics[width=0.96\textwidth]{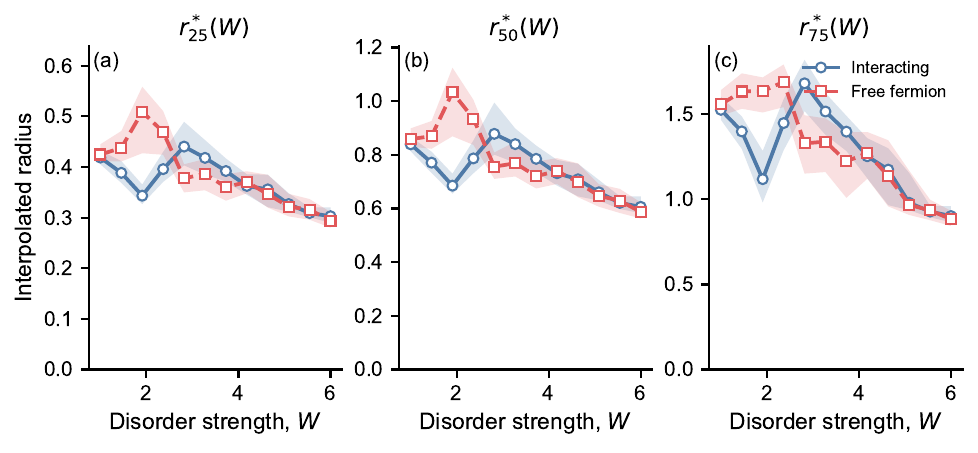}
    \caption{
        Threshold robustness of the recovery-radius proxy for the interacting kicked-Ising chain and the disordered free-fermion reference.
        The three plots show $r^\ast_{25}(W)$, $r^\ast_{50}(W)$, and $r^\ast_{75}(W)$ extracted from the same $N=12$ radius scans by varying the fraction of the total gain used to define the interpolated radius.
        Shaded bands denote bootstrap 68\% intervals over 100 disorder realizations.
    }
    \label{fig:rstar_thresholds}
\end{figure*}

The recovery-radius proxy is intentionally operational rather than microscopic, so its definition includes a threshold convention. Figure~\ref{fig:rstar_thresholds} checks the stability of the comparison by repeating the extraction for 25\%, 50\%, and 75\% of the total gain between $r=0$ and $r=3$. The absolute values shift monotonically with the threshold, as expected for an interpolated radius, but the qualitative distinction between the interacting and free-fermion scans remains intact: on the weak-to-crossover side, the free-fermion reference maintains a larger effective recovery length, consistent with its broader certified-gain profile, while both models approach comparable values toward strong disorder. We therefore treat $r^\ast_{50}(W)$ as a convenient summary proxy for the spatial scale of recoverable memory rather than as a uniquely preferred definition.

\section{LVR Implementation Details}
\label{app:lvr_details}

The local decoder is a layered circuit ansatz acting only on the recovery window $\Lambda_r(i)$.

\begin{equation}
\begin{aligned}
V_\theta
=&
\left[
\prod_{\ell=1}^{L}
\Big(
\prod_{j\in\Lambda_r(i)} \mathrm{Rot}_j(\theta_{\ell,j})
\Big)
\Big(
\prod_{(j,j+1)\subset \Lambda_r(i)} \mathrm{CZ}_{j,j+1}
\Big)
\right]
\\
&\times
\left(
\prod_{j\in\Lambda_r(i)} \mathrm{Rot}_j(\theta_{L+1,j})
\right),
\end{aligned}
\end{equation}

with $\mathrm{Rot}(\alpha,\beta,\gamma)=R_Z(\alpha)R_Y(\beta)R_Z(\gamma)$. For the main kicked-Ising model, we use $r=2$, so the decoder acts on a five-site window centered on the target qubit. The same ansatz family is used across all disorder strengths and pulse imperfections; only the variational angles are re-optimized for each disorder realization and control parameter.

For each input state $|\psi\rangle\in \mathcal{S}_6$, let the recovered reduced state of the target qubit be
\begin{equation}
\rho^{\mathrm{out}}_{i,\psi}
=
\frac{1}{2}
\left(
I+\mathbf{r}_\psi\cdot\boldsymbol{\sigma}
\right),
\end{equation}
with Bloch vector $\mathbf{r}_\psi$. If $\mathbf{n}_\psi$ denotes the unit Bloch vector of $|\psi\rangle$, then
\begin{equation}
\bra{\psi}\rho^{\mathrm{out}}_{i,\psi}\ket{\psi}
=
\frac{1+\mathbf{n}_\psi\cdot\mathbf{r}_\psi}{2}.
\end{equation}
For the six cardinal states, $\mathbf{n}_\psi\cdot\mathbf{r}_\psi$ is exactly
\(
\langle Z_i\rangle_{|0\rangle},
-\langle Z_i\rangle_{|1\rangle},
\langle X_i\rangle_{|+\rangle},
-\langle X_i\rangle_{|-\rangle},
\langle Y_i\rangle_{|+i\rangle},
-\langle Y_i\rangle_{|-i\rangle}
\),
so that the six-state average fidelity is
\begin{equation}
F_{\mathrm{avg},6}
=
\frac{1}{6}\sum_{\psi\in \mathcal{S}_6}
\bra{\psi}\rho^{\mathrm{out}}_{i,\psi}\ket{\psi}
=
\frac{1+Q_{\mathrm{LVR}}}{2}.
\label{eq:Favg6_relation}
\end{equation}
Equivalently, the six-state estimator used in the numerics is
\begin{equation}
Q_{\mathrm{LVR}}
=
\frac{1}{6}
\begin{aligned}[t]
\Big(&
\langle Z_i\rangle_{|0\rangle}
-\langle Z_i\rangle_{|1\rangle}
+\langle X_i\rangle_{|+\rangle}
-\langle X_i\rangle_{|-\rangle} \\
&+\langle Y_i\rangle_{|+i\rangle}
-\langle Y_i\rangle_{|-i\rangle}
\Big).
\end{aligned}
\end{equation}

\paragraph{Proof details for Identity~\ref{prop:sixstate_avg}.}
Let $\mathcal{N}$ be any recovered single-qubit channel with Bloch representation
\begin{equation}
\mathcal{N}\!\left[\frac{I+\mathbf{s}\cdot\boldsymbol{\sigma}}{2}\right]
=
\frac{I+(A\mathbf{s}+\mathbf{t})\cdot\boldsymbol{\sigma}}{2},
\label{eq:bloch_channel}
\end{equation}
where $A\in\mathbb R^{3\times 3}$ and $\mathbf t\in\mathbb R^3$. Define the six-state mean fidelity
\begin{equation}
F_6(\mathcal{N})
=
\frac{1}{6}\sum_{\psi\in\mathcal{S}_6}
\bra{\psi}\mathcal{N}\!\left(|\psi\rangle\langle\psi|\right)\ket{\psi}.
\label{eq:F6_def}
\end{equation}
Then
\begin{equation}
F_6(\mathcal{N})
=
\frac{1}{2}+\frac{\mathrm{Tr}(A)}{6}
=
F_{\mathrm{avg}}(\mathcal{N},\mathrm{id}),
\label{eq:F6_equals_Favg}
\end{equation}
and therefore
\begin{equation}
Q_{\mathrm{LVR}}
=
2F_{\mathrm{avg}}(\mathcal{N},\mathrm{id})-1.
\label{eq:QLVR_equals_channel_fidelity}
\end{equation}

\emph{Proof sketch.}
For a pure input state $|\psi\rangle$ with Bloch vector $\mathbf n_\psi$, the output fidelity is
\begin{equation}
\bra{\psi}\mathcal{N}\!\left(|\psi\rangle\langle\psi|\right)\ket{\psi}
=
\frac{1+\mathbf n_\psi\cdot(A\mathbf n_\psi+\mathbf t)}{2}.
\label{eq:pure_state_fidelity_bloch}
\end{equation}
The six cardinal Bloch vectors satisfy
\begin{equation}
\frac{1}{6}\sum_{\psi\in\mathcal{S}_6}\mathbf n_\psi=0,
\qquad
\frac{1}{6}\sum_{\psi\in\mathcal{S}_6}\mathbf n_\psi\mathbf n_\psi^{\mathsf T}=\frac{I_3}{3},
\label{eq:six_state_moments}
\end{equation}
so averaging Eq.~\eqref{eq:pure_state_fidelity_bloch} over $\mathcal{S}_6$ gives Eq.~\eqref{eq:F6_equals_Favg}. The uniform pure-state average over the Bloch sphere obeys the same first and second moments, so the Haar-average channel fidelity with the identity is the same quantity:
$F_{\mathrm{avg}}(\mathcal{N},\mathrm{id})=\tfrac12+\mathrm{Tr}(A)/6$.
For the recovered decoder channel, $F_6(\mathcal{N})$ is exactly the six-state average $F_{\mathrm{avg},6}$ in Eq.~\eqref{eq:Favg6_relation}, so Eq.~\eqref{eq:QLVR_equals_channel_fidelity} follows from Eq.~\eqref{eq:Favg6_relation}.

The recovery score depends, in principle, on the background product state $|\phi_{\mathrm{bg}}\rangle$ used on the non-target qubits, since that choice fixes the effective channel seen by the target qubit after the many-body evolution. We use the N\'eel background because it is simple, spatially structured, and experimentally natural. Changing the background can shift the absolute recovery loss, but it does not alter the operational definition of finite-window decoding.

The numerical optimization follows the same pattern in both physical models. For each disorder realization, we initialize all variational angles from an independent Gaussian distribution of width $0.1$, optimize the decoder with the Adam optimizer, and record the terminal value of the loss. In the kicked-Ising study, the reference calculation uses $D=30$ Floquet periods, $L=6$ decoder layers, 300 Adam steps, learning rate $0.03$, and 100 disorder realizations for each $(N,W)$ point. Throughout, all plotted points report disorder medians with IQR error bars. For the kicked-Ising benchmark, the gap-ratio calculation reuses exactly the same disorder configurations as the LVR optimization, ensuring a one-to-one comparison between the two diagnostics. In an experimental implementation, the same protocol can be paired with standard readout-calibration or measurement-error-mitigation workflows before forming the six-state estimator~\cite{PRXQuantum.2.040326}.

\section{Optimal On-Site CPTP Baseline: Proof Details}
\label{app:onsite_cptp}

The on-site baseline $Q^{\mathrm{opt}}_0$ is obtained by tracing the exact radius-$2$ reduced state down to the target qubit and then maximizing the same six-state score over all single-qubit CPTP maps. Concretely, for each disorder realization and each disorder strength, we construct the six target-site output states associated with the six probe inputs and solve a Choi-matrix semidefinite program over all maps from the target output qubit back to the target qubit. This produces the strongest possible one-site recovery benchmark compatible with the local output state itself. Because the optimization is over \emph{all} one-qubit CPTP maps, a positive
\begin{equation}
\Delta^{\mathrm{cert}}_r(\mathbf h)
=
Q^{\mathrm{shallow}}_r(\mathbf h)-Q^{\mathrm{opt}}_0(\mathbf h)
\end{equation}
cannot be attributed to a weak on-site ansatz. It certifies that recoverable information is present outside the target site but inside the chosen finite window.

Writing $\mathcal{E}_0$ for the channel from the target input qubit to the target-site output and $\mathcal{E}_r$ for the channel to the full radius-$r$ window, the two CPTP baselines can be written as
\begin{equation}
\begin{aligned}
Q^{\mathrm{opt}}_0
&=
\max_{\mathcal{D}_0\ \mathrm{CPTP}}
\Big[2F_6(\mathcal{D}_0\circ\mathcal{E}_0)-1\Big],
\\
Q^{\mathrm{opt}}_r
&=
\max_{\mathcal{D}_r\ \mathrm{CPTP}}
\Big[2F_6(\mathcal{D}_r\circ\mathcal{E}_r)-1\Big].
\end{aligned}
\label{eq:Qopt_decoder_defs}
\end{equation}

\paragraph{Proof details for Observation~\ref{prop:postprocess_no_gain}.}
If the radius-$r$ window output is only a CPTP post-processing of the target-site output,
\begin{equation}
\mathcal{E}_r
=
\mathcal{T}_r\circ \mathcal{E}_0,
\label{eq:postprocessing_assumption}
\end{equation}
then
\begin{equation}
Q^{\mathrm{opt}}_r=Q^{\mathrm{opt}}_0.
\label{eq:postprocessing_no_gain}
\end{equation}
Therefore any observed $Q^{\mathrm{shallow}}_r>Q^{\mathrm{opt}}_0$ certifies recoverable information outside the target site but within the chosen window.

\paragraph{Proof details for Observation~\ref{prop:postprocess_stability}.}
Assume that there exists a CPTP map $\mathcal{T}_r$ such that
\[
\|\mathcal{E}_r-\mathcal{T}_r\circ\mathcal{E}_0\|_\diamond \le \varepsilon.
\]
For any CPTP decoder $\mathcal{D}_r$ acting on the radius-$r$ window, define
\[
\mathcal{N}_1=\mathcal{D}_r\circ \mathcal{E}_r,
\qquad
\mathcal{N}_2=\mathcal{D}_r\circ \mathcal{T}_r\circ \mathcal{E}_0.
\]
Because $Q(\mathcal{N})=2F_6(\mathcal{N})-1$, and each fidelity term in $F_6$ is evaluated on a pure probe state, we have
\[
|Q(\mathcal{N}_1)-Q(\mathcal{N}_2)|
\le \|\mathcal{N}_1-\mathcal{N}_2\|_\diamond.
\]
Moreover, the diamond norm is contractive under CPTP post-processing, so
\[
\|\mathcal{N}_1-\mathcal{N}_2\|_\diamond
=
\|\mathcal{D}_r\circ \mathcal{E}_r-\mathcal{D}_r\circ \mathcal{T}_r\circ \mathcal{E}_0\|_\diamond
\le
\|\mathcal{E}_r-\mathcal{T}_r\circ \mathcal{E}_0\|_\diamond
\le \varepsilon.
\]
Therefore
\[
Q(\mathcal{D}_r\circ \mathcal{E}_r)
\le
Q(\mathcal{D}_r\circ \mathcal{T}_r\circ \mathcal{E}_0)+\varepsilon.
\]
Since $\mathcal{D}_r\circ \mathcal{T}_r$ is itself a CPTP map acting on the target-site output, the right-hand side is upper-bounded by $Q^{\mathrm{opt}}_0+\varepsilon$. Maximizing over all $\mathcal{D}_r$ yields
\[
Q^{\mathrm{opt}}_r \le Q^{\mathrm{opt}}_0+\varepsilon.
\]
Finally, because $Q^{\mathrm{shallow}}_r\le Q^{\mathrm{opt}}_r$, it follows that
\[
\Delta^{\mathrm{cert}}_r
=
Q^{\mathrm{shallow}}_r-Q^{\mathrm{opt}}_0
\le \varepsilon.
\]
Hence a measured certified gain larger than $\varepsilon$ excludes any $\varepsilon$-approximate target-site post-processing explanation of the window output.

\mainstatement{Observation}{Stability of the no-gain criterion}{prop:postprocess_stability}
If there exists a CPTP map $\mathcal{T}_r$ such that
\[
\|\mathcal{E}_r-\mathcal{T}_r\circ\mathcal{E}_0\|_\diamond \le \varepsilon,
\]
then
\[
Q^{\mathrm{opt}}_r \le Q^{\mathrm{opt}}_0 + \varepsilon,
\qquad
\Delta^{\mathrm{cert}}_r \le \varepsilon.
\]
Thus an observed certified gain $\Delta^{\mathrm{cert}}_r>\varepsilon$ rules out any description of the radius-$r$ window output as an $\varepsilon$-approximate CPTP post-processing of the target-site output.

\section{Paired-Disorder Statistics for the Certified Gain}
\label{app:paired_stats}

The descriptive summaries are computed from the paired raw disorder data used to define
\begin{equation}
\Delta^{\mathrm{cert}}_2(\mathbf h)
=
Q^{\mathrm{shallow}}_2(\mathbf h)-Q^{\mathrm{opt}}_0(\mathbf h).
\end{equation}
Here the radius-$2$ shallow scores and the on-site baselines are taken from matched numerical data sets that use the same 100 disorder realizations in the same order over the shared 12-point disorder grid.

For each disorder value $W$, we evaluate a one-sided paired Wilcoxon signed-rank test of the null hypothesis of zero paired difference against the alternative that $\Delta^{\mathrm{cert}}_2$ is positive, using the disorder realizations as the paired samples. We also form percentile bootstrap 95\% intervals for both the median of $\Delta^{\mathrm{cert}}_2$ and the positive fraction $\mathrm{Frac}(\Delta^{\mathrm{cert}}_2>0)$ by resampling the paired realizations with replacement $10^4$ times at each $W$.

Across the full scan, the pointwise Wilcoxon $p$ values range from $2.5\times 10^{-18}$ to $2.5\times 10^{-13}$. The bootstrap 95\% intervals for the median certified gain remain strictly above zero at every disorder point: their lower endpoints range from about $0.104$ to $0.254$, while their upper endpoints range from about $0.172$ to $0.407$. The corresponding bootstrap 95\% intervals for the positive-fraction curve remain high throughout, ranging from about $[0.77,0.91]$ at the broadest part of the scan to about $[0.95,1.00]$ near the strongest positive-fraction points. These pointwise statistics serve as paired-disorder robustness checks on the operational witness rather than as a separate model-selection procedure.

\section{Channel-Information Benchmark and Response-Operator Picture}
\label{app:channel_info}

The same six probe states used by LVR also suffice to reconstruct the full single-qubit input channel into the radius-$r$ window. Questions of informational uniqueness in tomography are separate from the present recovery benchmark and have been analyzed in optimization-based settings~\cite{WuAnZhangZhuHuangZeng2025}. Writing the target input as
\begin{equation}
\rho_{\mathrm{in}}
=
\frac{1}{2}\left(I+s_x X+s_y Y+s_z Z\right),
\end{equation}
the exact reduced output on $\Lambda_r$ can be written as
\begin{equation}
\mathcal{E}_r(\rho_{\mathrm{in}})
=
\rho_{\mathrm{av}}^{(r)}
+ s_x A_x^{(r)}
+ s_y A_y^{(r)}
+ s_z A_z^{(r)},
\end{equation}
with
\begin{equation}
\rho_{\mathrm{av}}^{(r)}=\frac{\rho_{|0\rangle}^{(r)}+\rho_{|1\rangle}^{(r)}}{2},
\qquad
A_\mu^{(r)}=
\frac{\rho_{|+\mu\rangle}^{(r)}-\rho_{|-\mu\rangle}^{(r)}}{2},
\end{equation}
where $\mu=x,y,z$ and $\rho_{|\psi\rangle}^{(r)}$ denotes the exact output state on the recovery window for probe $|\psi\rangle$.
The corresponding Choi state is
\begin{equation}
\Omega_{R\Lambda_r}
=
\frac{1}{2}
\begin{pmatrix}
\rho_{|0\rangle}^{(r)} & A_x^{(r)} + i A_y^{(r)} \\
A_x^{(r)} - i A_y^{(r)} & \rho_{|1\rangle}^{(r)}
\end{pmatrix},
\end{equation}
from which we evaluate the channel mutual information and coherent information,
\begin{equation}
I(R:\Lambda_r)=S(\Omega_R)+S(\Omega_{\Lambda_r})-S(\Omega_{R\Lambda_r}),
\end{equation}
\begin{equation}
I_{\mathrm{c}}(R\rangle \Lambda_r)=S(\Omega_{\Lambda_r})-S(\Omega_{R\Lambda_r}).
\end{equation}
Because the input is a single qubit and the channel is trace preserving, $S(\Omega_R)=1$ bit, so the site-to-window increments
\begin{equation}
\begin{aligned}
\Delta I_r
&= I(R:\Lambda_r)-I(R:\Lambda_0),
\\
\Delta I_{\mathrm{c},r}
&= I_{\mathrm{c}}(R\rangle \Lambda_r)-I_{\mathrm{c}}(R\rangle \Lambda_0)
\end{aligned}
\end{equation}
coincide.

This decomposition also clarifies what is and is not surprising about the operational hierarchy. For any fixed finite window, some intermediate-disorder enhancement over the on-site limit is expected: the gain is small when the response operators $A_\mu^{(r)}$ remain essentially on the target site and again when a substantial part of their support has spread beyond the chosen window. The nontrivial numerical statement is therefore not the bare existence of a peak, but that in the kicked-Ising chain the radius-$2$ shallow decoder remains well above the optimal on-site baseline while staying close to the same-window optimum across that intermediate window.

The same point can be phrased in operator-spreading language. Let
\begin{equation}
O_\mu(D)=U^\dagger(D)\,\sigma_i^\mu\,U(D)=\sum_S c^{(\mu)}_S(D,W)\,P_S
\end{equation}
be the Heisenberg-evolved Pauli operator on the target site, expanded in Pauli strings $P_S$ with support sets $S$. A finite-window recovery protocol can only exploit the part of this response whose support lies inside the chosen decoder window. If we define a cumulative local-response weight
\begin{equation}
p_r^{(\mu)}(D,W)=\sum_{S\subseteq \Lambda_r(i)} |c^{(\mu)}_S(D,W)|^2,
\end{equation}
then the on-site baseline is controlled by $p_0^{(\mu)}$, while the radius-$r$ window can only benefit from the shell weight between $p_0^{(\mu)}$ and $p_r^{(\mu)}$. This gives a heuristic Lieb-Robinson-style picture for the certified gain: it is small when the spreading length is still shorter than one site, becomes appreciable once the response has left the target but remains largely inside $\Lambda_r$, and falls again once a significant fraction of the operator weight has propagated beyond that window. The radius-scan proxy of Appendix \ref{app:rstar_thresholds} is the numerical counterpart of this picture.

\begin{figure*}[t]
    \centering
    \includegraphics[width=0.88\textwidth]{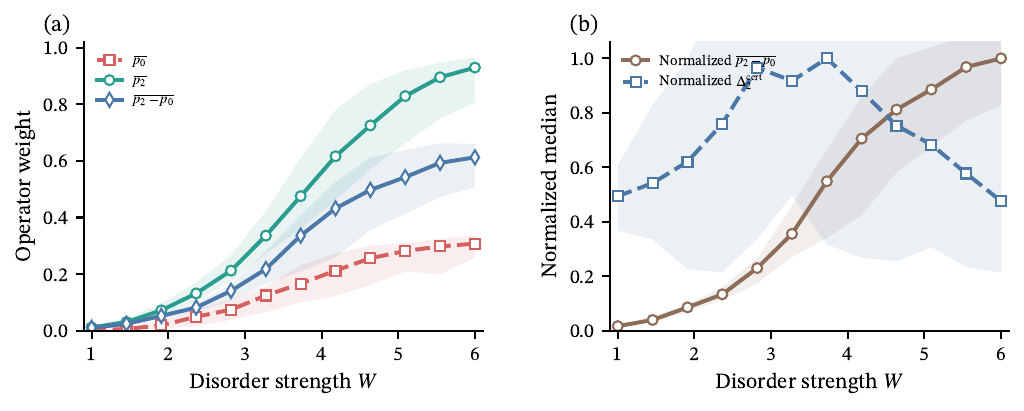}
    \caption{
        Operator shell weight tracks off-site support but not the full operational recoverability profile.
        (a) Exact Heisenberg shell weights averaged over $\mu=x,y,z$: the on-site weight $\overline{p_0}$, the five-site window weight $\overline{p_2}$, and their difference $\overline{p_2-p_0}$.
        (b) Normalized comparison between the exact shell weight and the certified gain $\Delta^{\mathrm{cert}}_2$.
        Curves denote disorder medians and shaded bands indicate the IQR.
    }
    \label{fig:operator_shell_weights}
\end{figure*}

Figure~\ref{fig:operator_shell_weights} provides a microscopic counterpart to the operational hierarchy. The exact shell weight $\overline{p_2-p_0}$ rises once the Heisenberg-evolved Pauli operators develop appreciable support away from the target but still inside the five-site window, so it captures the onset of off-site local memory. It does not, however, reproduce the operational profile: across the 36 paired realizations shown here, the shell-weight median continues rising all the way to the strongest disorder point, whereas $\Delta^{\mathrm{cert}}_2$ peaks near the crossover and falls to about half of that peak value by $W=6$. Local support inside the accessible window is therefore necessary but not sufficient for shallow target recovery. The missing ingredient is operational decodability: the shell weight counts all response retained somewhere inside $\Lambda_2$, while LVR asks which part of that content can actually be refocused onto the target qubit under the same locality and readout constraints as the protocol.

\section{Additional Background and Optimization Controls}
\label{app:controls}

This section collects two additional checks on background-state dependence and optimization reliability that support the same finite-window-recovery interpretation.
Throughout the appendix control plots, the recovery loss is reported as $\mathcal{L}=1-Q$, where $Q=2F_{\mathrm{avg}}-1$. Thus $\mathcal{L}=2(1-F_{\mathrm{avg}})$, and it should not be confused with the unrescaled infidelity $1-F_{\mathrm{avg}}$.

\begin{figure*}[t]
    \centering
    \includegraphics[width=0.88\textwidth]{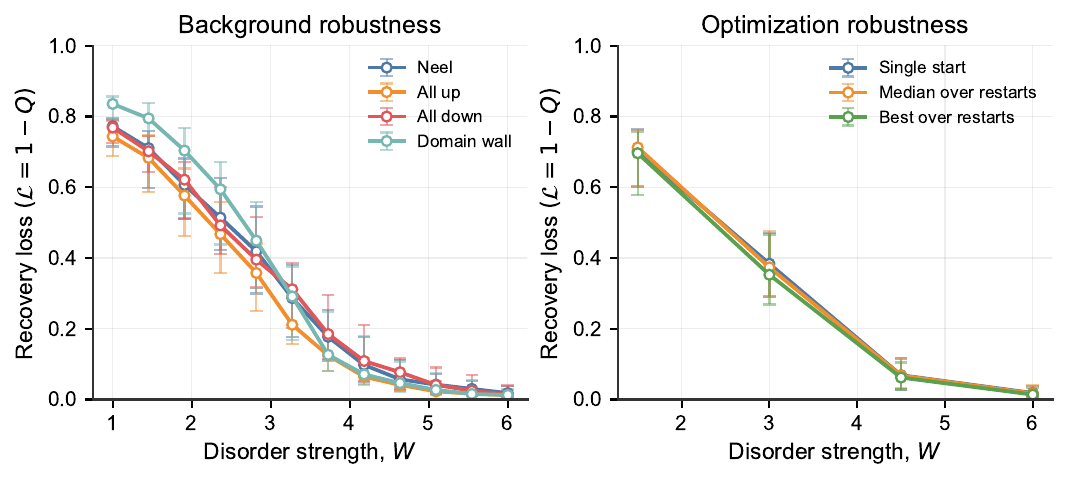}
    \caption{
        The disorder profile is not an artifact of a single background state or a single optimization restart.
        (a) Optimized recovery loss versus disorder strength $W$ for several product-state backgrounds on the non-target qubits.
        (b) Comparison between a single random initialization, the median terminal loss over 16 random restarts, and the best terminal loss over those restarts at representative disorder values.
        Markers denote disorder medians and error bars indicate the IQR over realizations.
    }
    \label{fig:controls_appendix}
\end{figure*}

Figure~\ref{fig:controls_appendix} collects the two most direct robustness checks on the optimized LVR signal. Figure~\ref{fig:controls_appendix}(a) shows that changing the product-state background shifts the absolute loss but preserves the ordering between weak-, intermediate-, and strong-disorder regimes. The crossover phenomenology is therefore not tied to a single specially chosen initialization. Figure~\ref{fig:controls_appendix}(b) addresses optimization reliability: across representative disorder strengths, the single-start, median-over-restarts, and best-over-restarts losses remain close, with only modest improvement from selecting the best restart. The disorder dependence is therefore unlikely to originate from isolated poor minima or systematic optimization failure in the crossover regime.

\section{Additional Same-Window and Expressivity Controls}
\label{app:cptp_control}

\clearpage
\onecolumngrid

\begin{figure}[t]
    \centering
    \includegraphics[width=0.70\textwidth]{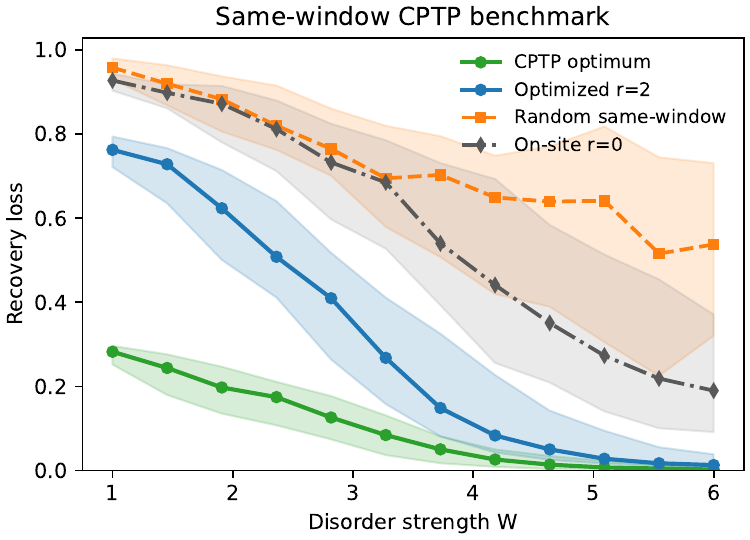}
    \caption{
        The shallow decoder remains below the optimal same-window CPTP benchmark while preserving the same crossover structure.
        The optimized shallow decoder is compared against the shallow on-site decoder, the best-of-32 random same-window circuits, and the optimal CPTP map acting on the same five-site window obtained from a Choi-matrix SDP.
        Markers denote medians and shaded bands indicate the IQR over realizations.
    }
    \label{fig:cptp_control}
\end{figure}

To separate ansatz limitations from the underlying finite-window information content, we also evaluate the optimal same-window CPTP benchmark across the full disorder scan. Starting from the exact reduced density matrix on the $r=2$ window, we optimize over all completely positive trace-preserving maps from the five-site window to the target qubit using a Choi-matrix semidefinite program. The quantity optimized by the SDP is the recoverable score $Q$, so it is an upper bound on $Q^{\mathrm{shallow}}_2$; equivalently, on the plotted loss axis $\mathcal{L}=1-Q$, it appears as a lower curve. Figure~\ref{fig:cptp_control} should therefore be read as
\begin{equation}
Q^{\mathrm{opt}}_2 \ge Q^{\mathrm{shallow}}_2,
\qquad
\mathcal{L}_2^{\mathrm{opt}} \le \mathcal{L}_2^{\mathrm{shallow}}.
\end{equation}
With that convention, the figure shows a clear hierarchy: the optimal same-window CPTP benchmark lies below the shallow radius-$2$ decoder on the loss axis, while the best random same-window circuits and the shallow on-site decoder remain noticeably worse. The matched random-circuit ensemble therefore rules out trivial same-window flexibility, whereas the CPTP optimization quantifies the full recoverable content permitted by the reduced window state itself.

This section collects auxiliary same-window random-circuit, decoder-depth, and finite-resource controls.

\begin{figure}[t]
    \centering
    \includegraphics[width=0.64\textwidth]{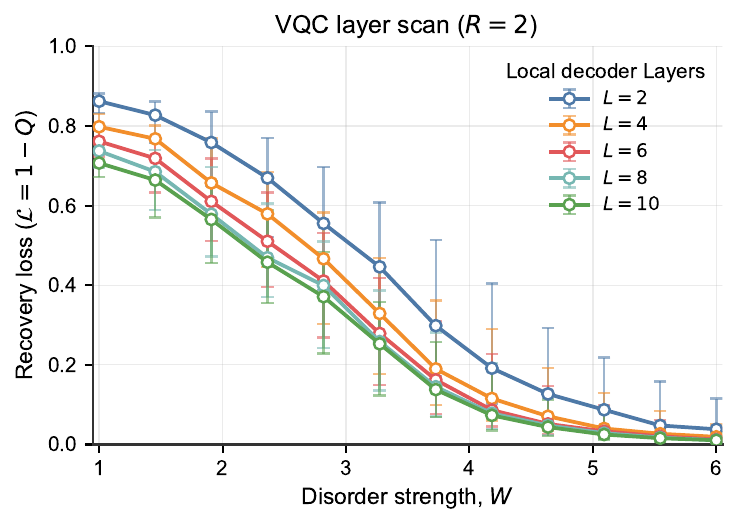}
    \caption{
        Decoder-layer robustness of the kicked-Ising recovery loss at fixed $N=12$, $r=2$, and $D=30$.
        The plotted curves compare optimized loss versus disorder strength for decoder depths $L=2,4,6,8,10$.
        The largest changes occur between shallow decoders and the $L=6$ ansatz used in the reference benchmark, while the deeper $L=8$ and $L=10$ curves produce only smaller additional changes.
    }
    \label{fig:layer_scan}
\end{figure}

Figure~\ref{fig:layer_scan} provides an expressivity control for the decoder choice. Increasing the decoder depth lowers the loss across the full disorder scan, as expected for a more expressive local ansatz, but the ordering between weak-, intermediate-, and strong-disorder regimes stabilizes once the decoder reaches modest depth. In particular, the crossover structure is already well developed by $L=6$, and the $L=8$ and $L=10$ curves remain close to the $L=6$ result over most of the scan. We therefore use $L=6$ as the reference depth, balancing decoder expressivity against circuit complexity rather than selecting a finely tuned point.

\begin{figure*}[t]
    \centering
    \includegraphics[width=0.96\textwidth]{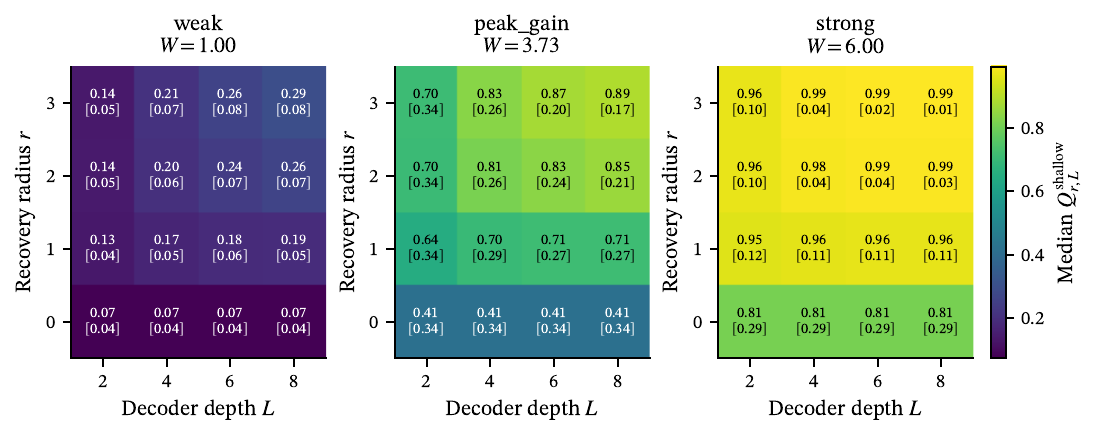}
    \caption{
        Finite-resource map of shallow finite-window recovery versus locality budget $r$ and decoder-depth budget $L$ on the exact $N=12$ kicked-Ising benchmark.
        The three maps show the median $Q^{\mathrm{shallow}}_{r,L}$ over 100 disorder realizations at weak disorder ($W=1.0$), near the peak median certified gain ($W=3.727273$), and strong disorder ($W=6.0$), scanning $r\in\{0,1,2,3\}$ and $L\in\{2,4,6,8\}$ at fixed $D=30$.
        The weak- and peak-disorder maps show diminishing but not fully saturated gains across the scanned grid, without a perfectly monotone improvement at every adjacent step, whereas the strong-disorder map is already close to saturation by about $r\approx 2$ and $L\approx 6$.
        This appendix figure is a fixed-benchmark resource-boundary illustration and does not support a system-size scaling claim or an asymptotic decoder-complexity law.
    }
    \label{fig:resource_map_rl}
\end{figure*}

Figure~\ref{fig:resource_map_rl} makes the finite-resource interpretation explicit on the exact $N=12$ benchmark by treating $r$ as the locality budget and $L$ as the decoder-expressivity budget. At weak disorder and near the certified-gain peak, increasing either budget still helps across the scanned grid, but the incremental gains already diminish and the median map is not perfectly monotone at every adjacent $(r,L)$ step. At strong disorder, by contrast, the median map is already close to saturated by about $r\approx 2$ and $L\approx 6$, so enlarging the window or deepening the decoder further produces only small additional changes over the scanned range. The figure should therefore be read only as a finite-resource boundary illustration for the fixed $N=12$, $D=30$ benchmark, not as evidence for system-size scaling or optimal resource exponents.

\section{Matched Evolution-Depth Control}
\label{app:dscan_control}

\clearpage
The fixed-window hierarchy in the main text is quoted at a representative depth $D=30$. Figure~\ref{fig:dscan_hierarchy} shows the matched depth scan used to check that this choice is not hiding a qualitatively different ordering at earlier or later times. The three cases isolate weak, crossover, and strong disorder so the depth dependence can be compared on the same disorder samples.

\refstepcounter{figure}
\begin{center}
    \includegraphics[width=0.80\textwidth]{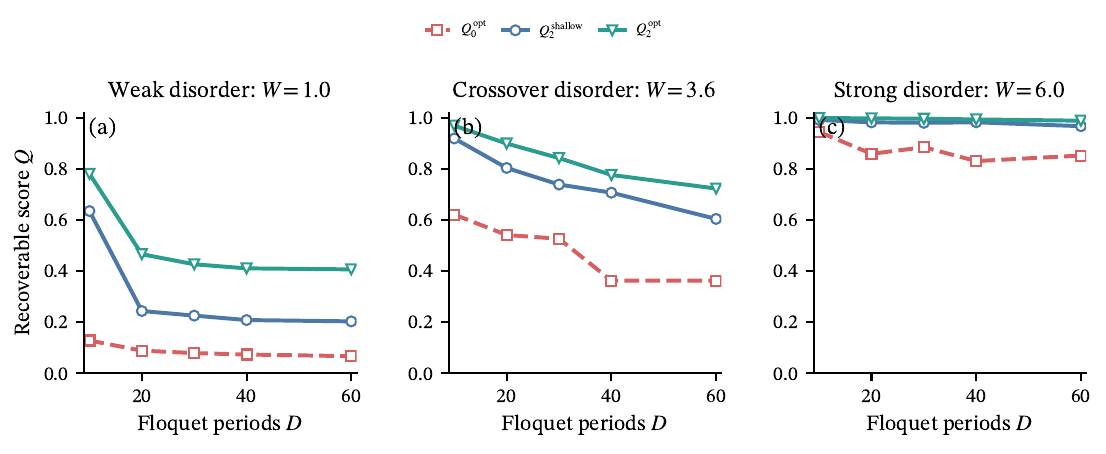}

    \parbox{0.88\textwidth}{\textbf{FIG.~\thefigure.} Matched evolution-depth scan of the recoverability hierarchy.
    The figure shows exact $D$-scan traces for the kicked-Ising chain at $N=12$ and $r=2$, evaluated on the same disorder ensemble at representative disorder strengths.
    (a) Weak disorder, (b) crossover disorder, and (c) strong disorder.}
    \label{fig:dscan_hierarchy}
\end{center}
\newpage

\section{Additional Radius and Time Dependence}
\label{app:extra_dynamics}

Figure~\ref{fig:extra_dynamics} extends the finite-window-gain comparison to a longer time axis and to two recovery radii. The purpose of this appendix is to verify that the operational gain remains nonzero over the scanned drive depths and that the qualitative time dependence is not restricted to the single depth used in the main benchmark. The free-fermion traces are included here only as an appendix-level control on how the time profile depends on the underlying transport mechanism.

\refstepcounter{figure}
\begin{center}
    \includegraphics[width=0.78\textwidth]{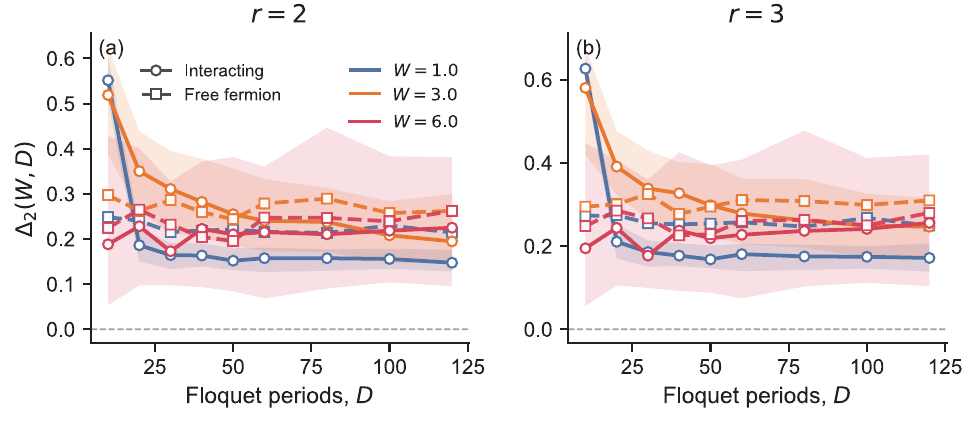}

    \parbox{0.88\textwidth}{\textbf{FIG.~\thefigure.} Time-dependence robustness check for the finite-window gain in the interacting kicked-Ising chain and in the disordered free-fermion reference.
    (a) and (b) show $\Delta_r(W,D)=\mathcal{L}^\star_{r=0}(W,D)-\mathcal{L}^\star_r(W,D)$ for $r=2$ and $r=3$, respectively, at representative disorder strengths $W=1,3,6$.
    Solid curves denote the interacting kicked-Ising data and dashed curves denote the free-fermion reference; the traces are shown as disorder medians, with uncertainty bands omitted here.
    Here $N=12$ and $L=6$ throughout, with medians taken over 100 disorder realizations.}
    \label{fig:extra_dynamics}
\end{center}
\newpage
\section{Geometry Effects and Boundary Conditions}
\label{app:geometry}

Figure~\ref{fig:geometry} checks how the local-recovery picture depends on boundary geometry. For a bulk target, the accessible window and its interface with the rest of the chain are nearly the same under periodic and open boundaries, so the loss curves should remain close. For an edge target, open boundaries reduce leakage channels out of the window and therefore provide a direct control on how recoverability depends on the local geometry of the inaccessible environment.

\refstepcounter{figure}
\begin{center}
    \includegraphics[width=0.42\textwidth,height=0.36\textheight,keepaspectratio]{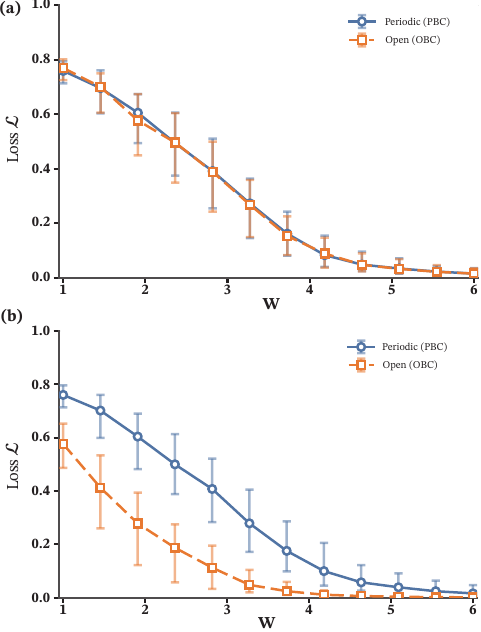}

    \parbox{0.88\textwidth}{\textbf{FIG.~\thefigure.} Boundary-geometry dependence of local recovery.
    Disorder-averaged recovery loss $\mathcal{L}$ versus disorder strength $W$ for system size $N=12$ under periodic boundary conditions (PBC) and open boundary conditions (OBC).
    (a) Recovery targeted at a bulk qubit.
    (b) Recovery targeted at an edge qubit.
    In both cases the decoder acts on a 5-qubit window ($r=2$) with depth $L=6$, and the evolution time is fixed at $D=30$ Floquet periods. Data points denote disorder medians with IQR error bars over 100 realizations.}
    \label{fig:geometry}
\end{center}
\clearpage
\twocolumngrid

\newpage


\begin{thebibliography}{46}%
\makeatletter
\providecommand \@ifxundefined [1]{%
 \@ifx{#1\undefined}
}%
\providecommand \@ifnum [1]{%
 \ifnum #1\expandafter \@firstoftwo
 \else \expandafter \@secondoftwo
 \fi
}%
\providecommand \@ifx [1]{%
 \ifx #1\expandafter \@firstoftwo
 \else \expandafter \@secondoftwo
 \fi
}%
\providecommand \natexlab [1]{#1}%
\providecommand \enquote  [1]{``#1''}%
\providecommand \bibnamefont  [1]{#1}%
\providecommand \bibfnamefont [1]{#1}%
\providecommand \citenamefont [1]{#1}%
\providecommand \href@noop [0]{\@secondoftwo}%
\providecommand \href [0]{\begingroup \@sanitize@url \@href}%
\providecommand \@href[1]{\@@startlink{#1}\@@href}%
\providecommand \@@href[1]{\endgroup#1\@@endlink}%
\providecommand \@sanitize@url [0]{\catcode `\\12\catcode `\$12\catcode
  `\&12\catcode `\#12\catcode `\^12\catcode `\_12\catcode `\%12\relax}%
\providecommand \@@startlink[1]{}%
\providecommand \@@endlink[0]{}%
\providecommand \url  [0]{\begingroup\@sanitize@url \@url }%
\providecommand \@url [1]{\endgroup\@href {#1}{\urlprefix }}%
\providecommand \urlprefix  [0]{URL }%
\providecommand \Eprint [0]{\href }%
\providecommand \doibase [0]{https://doi.org/}%
\providecommand \selectlanguage [0]{\@gobble}%
\providecommand \bibinfo  [0]{\@secondoftwo}%
\providecommand \bibfield  [0]{\@secondoftwo}%
\providecommand \translation [1]{[#1]}%
\providecommand \BibitemOpen [0]{}%
\providecommand \bibitemStop [0]{}%
\providecommand \bibitemNoStop [0]{.\EOS\space}%
\providecommand \EOS [0]{\spacefactor3000\relax}%
\providecommand \BibitemShut  [1]{\csname bibitem#1\endcsname}%
\let\auto@bib@innerbib\@empty
\bibitem [{\citenamefont {Gornyi}\ \emph {et~al.}(2005)\citenamefont {Gornyi},
  \citenamefont {Mirlin},\ and\ \citenamefont
  {Polyakov}}]{PhysRevLett.95.206603}%
  \BibitemOpen
  \bibfield  {author} {\bibinfo {author} {\bibfnamefont {I.~V.}\ \bibnamefont
  {Gornyi}}, \bibinfo {author} {\bibfnamefont {A.~D.}\ \bibnamefont {Mirlin}},\
  and\ \bibinfo {author} {\bibfnamefont {D.~G.}\ \bibnamefont {Polyakov}},\
  }\href {https://doi.org/10.1103/PhysRevLett.95.206603} {\bibfield  {journal}
  {\bibinfo  {journal} {Phys. Rev. Lett.}\ }\textbf {\bibinfo {volume} {95}},\
  \bibinfo {pages} {206603} (\bibinfo {year} {2005})}\BibitemShut {NoStop}%
\bibitem [{\citenamefont {Basko}\ \emph {et~al.}(2006)\citenamefont {Basko},
  \citenamefont {Aleiner},\ and\ \citenamefont {Altshuler}}]{Basko2006}%
  \BibitemOpen
  \bibfield  {author} {\bibinfo {author} {\bibfnamefont {D.}~\bibnamefont
  {Basko}}, \bibinfo {author} {\bibfnamefont {I.}~\bibnamefont {Aleiner}},\
  and\ \bibinfo {author} {\bibfnamefont {B.}~\bibnamefont {Altshuler}},\ }\href
  {https://doi.org/10.1016/j.aop.2005.11.014} {\bibfield
  {journal} {\bibinfo  {journal} {Annals of Physics}\ }\textbf {\bibinfo
  {volume} {321}},\ \bibinfo {pages} {1126} (\bibinfo {year}
  {2006})}\BibitemShut {NoStop}%
\bibitem [{\citenamefont {\ifmmode \check{Z}\else
  \v{Z}\fi{}nidari\ifmmode~\check{c}\else \v{c}\fi{}}\ \emph
  {et~al.}(2008)\citenamefont {\ifmmode \check{Z}\else
  \v{Z}\fi{}nidari\ifmmode~\check{c}\else \v{c}\fi{}}, \citenamefont {Prosen},\
  and\ \citenamefont {Prelov\ifmmode~\check{s}\else
  \v{s}\fi{}ek}}]{PhysRevB.77.064426}%
  \BibitemOpen
  \bibfield  {author} {\bibinfo {author} {\bibfnamefont {M.}~\bibnamefont
  {\ifmmode \check{Z}\else \v{Z}\fi{}nidari\ifmmode~\check{c}\else
  \v{c}\fi{}}}, \bibinfo {author} {\bibfnamefont {T.~c.~v.}\ \bibnamefont
  {Prosen}},\ and\ \bibinfo {author} {\bibfnamefont {P.}~\bibnamefont
  {Prelov\ifmmode~\check{s}\else \v{s}\fi{}ek}},\ }\href
  {https://doi.org/10.1103/PhysRevB.77.064426} {\bibfield  {journal} {\bibinfo
  {journal} {Phys. Rev. B}\ }\textbf {\bibinfo {volume} {77}},\ \bibinfo
  {pages} {064426} (\bibinfo {year} {2008})}\BibitemShut {NoStop}%
\bibitem [{\citenamefont {Pal}\ and\ \citenamefont
  {Huse}(2010)}]{PhysRevB.82.174411}%
  \BibitemOpen
  \bibfield  {author} {\bibinfo {author} {\bibfnamefont {A.}~\bibnamefont
  {Pal}}\ and\ \bibinfo {author} {\bibfnamefont {D.~A.}\ \bibnamefont {Huse}},\
  }\href {https://doi.org/10.1103/PhysRevB.82.174411} {\bibfield  {journal}
  {\bibinfo  {journal} {Phys. Rev. B}\ }\textbf {\bibinfo {volume} {82}},\
  \bibinfo {pages} {174411} (\bibinfo {year} {2010})}\BibitemShut {NoStop}%
\bibitem [{\citenamefont {Nandkishore}\ and\ \citenamefont
  {Huse}(2015)}]{Nandkishore2015}%
  \BibitemOpen
  \bibfield  {author} {\bibinfo {author} {\bibfnamefont {R.}~\bibnamefont
  {Nandkishore}}\ and\ \bibinfo {author} {\bibfnamefont {D.~A.}\ \bibnamefont
  {Huse}},\ }\href
  {https://doi.org/10.1146/annurev-conmatphys-031214-014726}
  {\bibfield  {journal} {\bibinfo  {journal} {Annual Review of Condensed Matter
  Physics}\ }\textbf {\bibinfo {volume} {6}},\ \bibinfo {pages} {15} (\bibinfo
  {year} {2015})}\BibitemShut {NoStop}%
\bibitem [{\citenamefont {Alet}\ and\ \citenamefont
  {Laflorencie}(2018)}]{Alet2018}%
  \BibitemOpen
  \bibfield  {author} {\bibinfo {author} {\bibfnamefont {F.}~\bibnamefont
  {Alet}}\ and\ \bibinfo {author} {\bibfnamefont {N.}~\bibnamefont
  {Laflorencie}},\ }\href
  {https://doi.org/10.1016/j.crhy.2018.03.003} {\bibfield
  {journal} {\bibinfo  {journal} {Comptes Rendus Physique}\ }\textbf {\bibinfo
  {volume} {19}},\ \bibinfo {pages} {498} (\bibinfo {year} {2018})},\ \bibinfo
  {note} {quantum simulation / Simulation quantique}\BibitemShut {NoStop}%
\bibitem [{\citenamefont {Abanin}\ \emph {et~al.}(2019)\citenamefont {Abanin},
  \citenamefont {Altman}, \citenamefont {Bloch},\ and\ \citenamefont
  {Serbyn}}]{RevModPhys.91.021001}%
  \BibitemOpen
  \bibfield  {author} {\bibinfo {author} {\bibfnamefont {D.~A.}\ \bibnamefont
  {Abanin}}, \bibinfo {author} {\bibfnamefont {E.}~\bibnamefont {Altman}},
  \bibinfo {author} {\bibfnamefont {I.}~\bibnamefont {Bloch}},\ and\ \bibinfo
  {author} {\bibfnamefont {M.}~\bibnamefont {Serbyn}},\ }\href
  {https://doi.org/10.1103/RevModPhys.91.021001} {\bibfield  {journal}
  {\bibinfo  {journal} {Rev. Mod. Phys.}\ }\textbf {\bibinfo {volume} {91}},\
  \bibinfo {pages} {021001} (\bibinfo {year} {2019})}\BibitemShut {NoStop}%
\bibitem [{\citenamefont {Preskill}(2018)}]{Preskill2018quantumcomputingin}%
  \BibitemOpen
  \bibfield  {author} {\bibinfo {author} {\bibfnamefont {J.}~\bibnamefont
  {Preskill}},\ }\href {https://doi.org/10.22331/q-2018-08-06-79} {\bibfield
  {journal} {\bibinfo  {journal} {{Quantum}}\ }\textbf {\bibinfo {volume}
  {2}},\ \bibinfo {pages} {79} (\bibinfo {year} {2018})}\BibitemShut {NoStop}%
\bibitem [{\citenamefont {Lieb}\ and\ \citenamefont
  {Robinson}(1972)}]{Lieb1972}%
  \BibitemOpen
  \bibfield  {author} {\bibinfo {author} {\bibfnamefont {E.~H.}\ \bibnamefont
  {Lieb}}\ and\ \bibinfo {author} {\bibfnamefont {D.~W.}\ \bibnamefont
  {Robinson}},\ }\href {https://doi.org/10.1007/BF01645779} {\bibfield
  {journal} {\bibinfo  {journal} {Communications in Mathematical Physics}\
  }\textbf {\bibinfo {volume} {28}},\ \bibinfo {pages} {251} (\bibinfo {year}
  {1972})}\BibitemShut {NoStop}%
\bibitem [{\citenamefont {Nahum}\ \emph {et~al.}(2017)\citenamefont {Nahum},
  \citenamefont {Ruhman}, \citenamefont {Vijay},\ and\ \citenamefont
  {Haah}}]{Nahum2017}%
  \BibitemOpen
  \bibfield  {author} {\bibinfo {author} {\bibfnamefont {A.}~\bibnamefont
  {Nahum}}, \bibinfo {author} {\bibfnamefont {J.}~\bibnamefont {Ruhman}},
  \bibinfo {author} {\bibfnamefont {S.}~\bibnamefont {Vijay}},\ and\ \bibinfo
  {author} {\bibfnamefont {J.}~\bibnamefont {Haah}},\ }\href
  {https://doi.org/10.1103/PhysRevX.7.031016} {\bibfield  {journal} {\bibinfo
  {journal} {Physical Review X}\ }\textbf {\bibinfo {volume} {7}},\ \bibinfo
  {pages} {031016} (\bibinfo {year} {2017})}\BibitemShut {NoStop}%
\bibitem [{\citenamefont {Oganesyan}\ and\ \citenamefont
  {Huse}(2007)}]{PhysRevB.75.155111}%
  \BibitemOpen
  \bibfield  {author} {\bibinfo {author} {\bibfnamefont {V.}~\bibnamefont
  {Oganesyan}}\ and\ \bibinfo {author} {\bibfnamefont {D.~A.}\ \bibnamefont
  {Huse}},\ }\href {https://doi.org/10.1103/PhysRevB.75.155111} {\bibfield
  {journal} {\bibinfo  {journal} {Phys. Rev. B}\ }\textbf {\bibinfo {volume}
  {75}},\ \bibinfo {pages} {155111} (\bibinfo {year} {2007})}\BibitemShut
  {NoStop}%
\bibitem [{\citenamefont {Schreiber}\ \emph {et~al.}(2015)\citenamefont
  {Schreiber}, \citenamefont {Hodgman}, \citenamefont {Bordia}, \citenamefont
  {L{\"u}schen}, \citenamefont {Fischer}, \citenamefont {Vosk}, \citenamefont
  {Altman}, \citenamefont {Schneider},\ and\ \citenamefont
  {Bloch}}]{Schreiber2015}%
  \BibitemOpen
  \bibfield  {author} {\bibinfo {author} {\bibfnamefont {M.}~\bibnamefont
  {Schreiber}}, \bibinfo {author} {\bibfnamefont {S.~S.}\ \bibnamefont
  {Hodgman}}, \bibinfo {author} {\bibfnamefont {P.}~\bibnamefont {Bordia}},
  \bibinfo {author} {\bibfnamefont {H.~P.}\ \bibnamefont {L{\"u}schen}},
  \bibinfo {author} {\bibfnamefont {M.~H.}\ \bibnamefont {Fischer}}, \bibinfo
  {author} {\bibfnamefont {R.}~\bibnamefont {Vosk}}, \bibinfo {author}
  {\bibfnamefont {E.}~\bibnamefont {Altman}}, \bibinfo {author} {\bibfnamefont
  {U.}~\bibnamefont {Schneider}},\ and\ \bibinfo {author} {\bibfnamefont
  {I.}~\bibnamefont {Bloch}},\ }\href {https://doi.org/10.1126/science.aaa7432}
  {\bibfield  {journal} {\bibinfo  {journal} {Science}\ }\textbf {\bibinfo
  {volume} {349}},\ \bibinfo {pages} {842 } (\bibinfo {year}
  {2015})}\BibitemShut {NoStop}%
\bibitem [{\citenamefont {Smith}\ \emph {et~al.}(2016)\citenamefont {Smith},
  \citenamefont {Lee}, \citenamefont {Richerme}, \citenamefont {Neyenhuis},
  \citenamefont {Hess}, \citenamefont {Hauke}, \citenamefont {Heyl},
  \citenamefont {Huse},\ and\ \citenamefont {Monroe}}]{Smith2016}%
  \BibitemOpen
  \bibfield  {author} {\bibinfo {author} {\bibfnamefont {J.}~\bibnamefont
  {Smith}}, \bibinfo {author} {\bibfnamefont {A.}~\bibnamefont {Lee}}, \bibinfo
  {author} {\bibfnamefont {P.}~\bibnamefont {Richerme}}, \bibinfo {author}
  {\bibfnamefont {B.}~\bibnamefont {Neyenhuis}}, \bibinfo {author}
  {\bibfnamefont {P.~W.}\ \bibnamefont {Hess}}, \bibinfo {author}
  {\bibfnamefont {P.}~\bibnamefont {Hauke}}, \bibinfo {author} {\bibfnamefont
  {M.}~\bibnamefont {Heyl}}, \bibinfo {author} {\bibfnamefont {D.~A.}\
  \bibnamefont {Huse}},\ and\ \bibinfo {author} {\bibfnamefont
  {C.}~\bibnamefont {Monroe}},\ }\href {https://doi.org/10.1038/nphys3783}
  {\bibfield  {journal} {\bibinfo  {journal} {Nature Physics}\ }\textbf
  {\bibinfo {volume} {12}},\ \bibinfo {pages} {907} (\bibinfo {year}
  {2016})}\BibitemShut {NoStop}%
\bibitem [{\citenamefont {Xu}\ \emph {et~al.}(2018)\citenamefont {Xu},
  \citenamefont {Chen}, \citenamefont {Zeng}, \citenamefont {Zhang},
  \citenamefont {Song}, \citenamefont {Liu}, \citenamefont {Guo}, \citenamefont
  {Zhang}, \citenamefont {Xu}, \citenamefont {Deng}, \citenamefont {Huang},
  \citenamefont {Wang}, \citenamefont {Zhu}, \citenamefont {Zheng},\ and\
  \citenamefont {Fan}}]{PhysRevLett.120.050507}%
  \BibitemOpen
  \bibfield  {author} {\bibinfo {author} {\bibfnamefont {K.}~\bibnamefont
  {Xu}}, \bibinfo {author} {\bibfnamefont {J.-J.}\ \bibnamefont {Chen}},
  \bibinfo {author} {\bibfnamefont {Y.}~\bibnamefont {Zeng}}, \bibinfo {author}
  {\bibfnamefont {Y.-R.}\ \bibnamefont {Zhang}}, \bibinfo {author}
  {\bibfnamefont {C.}~\bibnamefont {Song}}, \bibinfo {author} {\bibfnamefont
  {W.}~\bibnamefont {Liu}}, \bibinfo {author} {\bibfnamefont {Q.}~\bibnamefont
  {Guo}}, \bibinfo {author} {\bibfnamefont {P.}~\bibnamefont {Zhang}}, \bibinfo
  {author} {\bibfnamefont {D.}~\bibnamefont {Xu}}, \bibinfo {author}
  {\bibfnamefont {H.}~\bibnamefont {Deng}}, \bibinfo {author} {\bibfnamefont
  {K.}~\bibnamefont {Huang}}, \bibinfo {author} {\bibfnamefont
  {H.}~\bibnamefont {Wang}}, \bibinfo {author} {\bibfnamefont {X.}~\bibnamefont
  {Zhu}}, \bibinfo {author} {\bibfnamefont {D.}~\bibnamefont {Zheng}},\ and\
  \bibinfo {author} {\bibfnamefont {H.}~\bibnamefont {Fan}},\ }\href
  {https://doi.org/10.1103/PhysRevLett.120.050507} {\bibfield  {journal}
  {\bibinfo  {journal} {Phys. Rev. Lett.}\ }\textbf {\bibinfo {volume} {120}},\
  \bibinfo {pages} {050507} (\bibinfo {year} {2018})}\BibitemShut {NoStop}%
\bibitem [{\citenamefont {Prosen}(2002)}]{PhysRevE.65.036208}%
  \BibitemOpen
  \bibfield  {author} {\bibinfo {author} {\bibfnamefont {T.~c.~v.}\
  \bibnamefont {Prosen}},\ }\href {https://doi.org/10.1103/PhysRevE.65.036208}
  {\bibfield  {journal} {\bibinfo  {journal} {Phys. Rev. E}\ }\textbf {\bibinfo
  {volume} {65}},\ \bibinfo {pages} {036208} (\bibinfo {year}
  {2002})}\BibitemShut {NoStop}%
\bibitem [{\citenamefont {Serbyn}\ and\ \citenamefont
  {Abanin}(2017)}]{PhysRevB.96.014202}%
  \BibitemOpen
  \bibfield  {author} {\bibinfo {author} {\bibfnamefont {M.}~\bibnamefont
  {Serbyn}}\ and\ \bibinfo {author} {\bibfnamefont {D.~A.}\ \bibnamefont
  {Abanin}},\ }\href {https://doi.org/10.1103/PhysRevB.96.014202} {\bibfield
  {journal} {\bibinfo  {journal} {Phys. Rev. B}\ }\textbf {\bibinfo {volume}
  {96}},\ \bibinfo {pages} {014202} (\bibinfo {year} {2017})}\BibitemShut
  {NoStop}%
\bibitem [{\citenamefont {Prosen}(2007)}]{Prosen2007}%
  \BibitemOpen
  \bibfield  {author} {\bibinfo {author} {\bibfnamefont {T.}~\bibnamefont
  {Prosen}},\ }\href {https://doi.org/10.1088/1751-8113/40/28/S02} {\bibfield
  {journal} {\bibinfo  {journal} {Journal of Physics A: Mathematical and
  Theoretical}\ }\textbf {\bibinfo {volume} {40}},\ \bibinfo {pages} {7881}
  (\bibinfo {year} {2007})}\BibitemShut {NoStop}%
\bibitem [{\citenamefont {Gärttner}\ \emph {et~al.}(2017)\citenamefont
  {Gärttner}, \citenamefont {Bohnet}, \citenamefont {Safavi-Naini},
  \citenamefont {Wall}, \citenamefont {Bollinger},\ and\ \citenamefont
  {Rey}}]{Gaerttner2017}%
  \BibitemOpen
  \bibfield  {author} {\bibinfo {author} {\bibfnamefont {M.}~\bibnamefont
  {Gärttner}}, \bibinfo {author} {\bibfnamefont {J.~G.}\ \bibnamefont
  {Bohnet}}, \bibinfo {author} {\bibfnamefont {A.}~\bibnamefont
  {Safavi-Naini}}, \bibinfo {author} {\bibfnamefont {M.~L.}\ \bibnamefont
  {Wall}}, \bibinfo {author} {\bibfnamefont {J.~J.}\ \bibnamefont
  {Bollinger}},\ and\ \bibinfo {author} {\bibfnamefont {A.~M.}\ \bibnamefont
  {Rey}},\ }\href {https://doi.org/10.1038/nphys4119} {\bibfield  {journal}
  {\bibinfo  {journal} {Nature Physics}\ }\textbf {\bibinfo {volume} {13}},\
  \bibinfo {pages} {781} (\bibinfo {year} {2017})}\BibitemShut {NoStop}%
\bibitem [{\citenamefont {García-Mata}\ \emph {et~al.}(2023)\citenamefont
  {García-Mata}, \citenamefont {Jalabert},\ and\ \citenamefont
  {Wisniacki}}]{García-Mata:2023}%
  \BibitemOpen
  \bibfield  {author} {\bibinfo {author} {\bibfnamefont {I.}~\bibnamefont
  {García-Mata}}, \bibinfo {author} {\bibfnamefont {R.~A.}\ \bibnamefont
  {Jalabert}},\ and\ \bibinfo {author} {\bibfnamefont {D.~A.}\ \bibnamefont
  {Wisniacki}},\ }\href {https://doi.org/10.4249/scholarpedia.55237} {\bibfield
   {journal} {\bibinfo  {journal} {Scholarpedia}\ }\textbf {\bibinfo {volume}
  {18}},\ \bibinfo {pages} {55237} (\bibinfo {year} {2023})},\ \bibinfo {note}
  {revision \#204529}\BibitemShut {NoStop}%
\bibitem [{\citenamefont {Gross}\ \emph {et~al.}(2010)\citenamefont {Gross},
  \citenamefont {Liu}, \citenamefont {Flammia}, \citenamefont {Becker},\ and\
  \citenamefont {Eisert}}]{Gross2010QST}%
  \BibitemOpen
  \bibfield  {author} {\bibinfo {author} {\bibfnamefont {D.}~\bibnamefont
  {Gross}}, \bibinfo {author} {\bibfnamefont {Y.-K.}\ \bibnamefont {Liu}},
  \bibinfo {author} {\bibfnamefont {S.~T.}\ \bibnamefont {Flammia}}, \bibinfo
  {author} {\bibfnamefont {S.}~\bibnamefont {Becker}},\ and\ \bibinfo {author}
  {\bibfnamefont {J.}~\bibnamefont {Eisert}},\ }\href
  {https://doi.org/10.1103/PhysRevLett.105.150401} {\bibfield  {journal}
  {\bibinfo  {journal} {Phys. Rev. Lett.}\ }\textbf {\bibinfo {volume} {105}},\
  \bibinfo {pages} {150401} (\bibinfo {year} {2010})}\BibitemShut {NoStop}%
\bibitem [{\citenamefont {An}\ \emph {et~al.}(2024)\citenamefont {An},
  \citenamefont {Wu}, \citenamefont {Yang}, \citenamefont {Zhou},\ and\
  \citenamefont {Zeng}}]{AnWuYangZhouZeng2024}%
  \BibitemOpen
  \bibfield  {author} {\bibinfo {author} {\bibfnamefont {Z.}~\bibnamefont
  {An}}, \bibinfo {author} {\bibfnamefont {J.}~\bibnamefont {Wu}}, \bibinfo
  {author} {\bibfnamefont {M.}~\bibnamefont {Yang}}, \bibinfo {author}
  {\bibfnamefont {D.~L.}\ \bibnamefont {Zhou}},\ and\ \bibinfo {author}
  {\bibfnamefont {B.}~\bibnamefont {Zeng}},\ }\href
  {https://doi.org/10.1103/PhysRevApplied.21.014037} {\bibfield  {journal}
  {\bibinfo  {journal} {Phys. Rev. Applied}\ }\textbf {\bibinfo {volume} {21}},\
  \bibinfo {pages} {014037} (\bibinfo {year} {2024})}\BibitemShut {NoStop}%
\bibitem [{\citenamefont {Barnum}\ and\ \citenamefont
  {Knill}(2002)}]{BarnumKnill2002}%
  \BibitemOpen
  \bibfield  {author} {\bibinfo {author} {\bibfnamefont {H.}~\bibnamefont
  {Barnum}}\ and\ \bibinfo {author} {\bibfnamefont {E.}~\bibnamefont {Knill}},\
  }\href {https://doi.org/10.1063/1.1459754} {\bibfield  {journal} {\bibinfo
  {journal} {Journal of Mathematical Physics}\ }\textbf {\bibinfo {volume}
  {43}},\ \bibinfo {pages} {2097} (\bibinfo {year} {2002})}\BibitemShut
  {NoStop}%
\bibitem [{\citenamefont {B{\'e}ny}\ and\ \citenamefont
  {Oreshkov}(2010)}]{BenyOreshkov2010}%
  \BibitemOpen
  \bibfield  {author} {\bibinfo {author} {\bibfnamefont {C.}~\bibnamefont
  {B{\'e}ny}}\ and\ \bibinfo {author} {\bibfnamefont {O.}~\bibnamefont
  {Oreshkov}},\ }\href {https://doi.org/10.1103/PhysRevLett.104.120501}
  {\bibfield  {journal} {\bibinfo  {journal} {Physical Review Letters}\
  }\textbf {\bibinfo {volume} {104}},\ \bibinfo {pages} {120501} (\bibinfo
  {year} {2010})}\BibitemShut {NoStop}%
\bibitem [{\citenamefont {Fawzi}\ and\ \citenamefont
  {Renner}(2015)}]{FawziRenner2015}%
  \BibitemOpen
  \bibfield  {author} {\bibinfo {author} {\bibfnamefont {O.}~\bibnamefont
  {Fawzi}}\ and\ \bibinfo {author} {\bibfnamefont {R.}~\bibnamefont {Renner}},\
  }\href {https://doi.org/10.1007/s00220-015-2466-x} {\bibfield  {journal}
  {\bibinfo  {journal} {Communications in Mathematical Physics}\ }\textbf
  {\bibinfo {volume} {340}},\ \bibinfo {pages} {575} (\bibinfo {year}
  {2015})}\BibitemShut {NoStop}%
\bibitem [{\citenamefont {Sutter}\ \emph {et~al.}(2016)\citenamefont {Sutter},
  \citenamefont {Fawzi},\ and\ \citenamefont {Renner}}]{Sutter2016}%
  \BibitemOpen
  \bibfield  {author} {\bibinfo {author} {\bibfnamefont {D.}~\bibnamefont
  {Sutter}}, \bibinfo {author} {\bibfnamefont {O.}~\bibnamefont {Fawzi}},\ and\
  \bibinfo {author} {\bibfnamefont {R.}~\bibnamefont {Renner}},\ }\href
  {https://doi.org/10.1098/rspa.2015.0623} {\bibfield  {journal} {\bibinfo
  {journal} {Proceedings of the Royal Society A}\ }\textbf {\bibinfo {volume}
  {472}},\ \bibinfo {pages} {20150623} (\bibinfo {year} {2016})}\BibitemShut
  {NoStop}%
\bibitem [{\citenamefont {An}\ \emph {et~al.}(2024)\citenamefont {An},
  \citenamefont {Cao}, \citenamefont {Xu},\ and\ \citenamefont
  {Zhou}}]{AnCaoXuZhou2024}%
  \BibitemOpen
  \bibfield  {author} {\bibinfo {author} {\bibfnamefont {Z.}~\bibnamefont
  {An}}, \bibinfo {author} {\bibfnamefont {C.}~\bibnamefont {Cao}}, \bibinfo
  {author} {\bibfnamefont {C.~Q.}\ \bibnamefont {Xu}},\ and\ \bibinfo {author}
  {\bibfnamefont {D.~L.}\ \bibnamefont {Zhou}},\ }\href
  {https://doi.org/10.22331/q-2024-07-22-1421} {\bibfield  {journal}
  {\bibinfo  {journal} {Quantum}\ }\textbf {\bibinfo {volume} {8}},\ \bibinfo
  {pages} {1421} (\bibinfo {year} {2024})}\BibitemShut {NoStop}%
\bibitem [{\citenamefont {Massar}\ and\ \citenamefont
  {Popescu}(1995)}]{MassarPopescu1995}%
  \BibitemOpen
  \bibfield  {author} {\bibinfo {author} {\bibfnamefont {S.}~\bibnamefont
  {Massar}}\ and\ \bibinfo {author} {\bibfnamefont {S.}~\bibnamefont
  {Popescu}},\ }\href {https://doi.org/10.1103/PhysRevLett.74.1259}
  {\bibfield  {journal} {\bibinfo  {journal} {Phys. Rev. Lett.}\ }\textbf
  {\bibinfo {volume} {74}},\ \bibinfo {pages} {1259} (\bibinfo {year}
  {1995})}\BibitemShut {NoStop}%
\bibitem [{\citenamefont {Ponte}\ \emph {et~al.}(2015)\citenamefont {Ponte},
  \citenamefont {Papi\ifmmode~\acute{c}\else \'{c}\fi{}}, \citenamefont
  {Huveneers},\ and\ \citenamefont {Abanin}}]{PhysRevLett.114.140401}%
  \BibitemOpen
  \bibfield  {author} {\bibinfo {author} {\bibfnamefont {P.}~\bibnamefont
  {Ponte}}, \bibinfo {author} {\bibfnamefont {Z.}~\bibnamefont
  {Papi\ifmmode~\acute{c}\else \'{c}\fi{}}}, \bibinfo {author} {\bibfnamefont
  {F.~m.~c.}\ \bibnamefont {Huveneers}},\ and\ \bibinfo {author} {\bibfnamefont
  {D.~A.}\ \bibnamefont {Abanin}},\ }\href
  {https://doi.org/10.1103/PhysRevLett.114.140401} {\bibfield  {journal}
  {\bibinfo  {journal} {Phys. Rev. Lett.}\ }\textbf {\bibinfo {volume} {114}},\
  \bibinfo {pages} {140401} (\bibinfo {year} {2015})}\BibitemShut {NoStop}%
\bibitem [{\citenamefont {Lazarides}\ \emph {et~al.}(2015)\citenamefont
  {Lazarides}, \citenamefont {Das},\ and\ \citenamefont
  {Moessner}}]{PhysRevLett.115.030402}%
  \BibitemOpen
  \bibfield  {author} {\bibinfo {author} {\bibfnamefont {A.}~\bibnamefont
  {Lazarides}}, \bibinfo {author} {\bibfnamefont {A.}~\bibnamefont {Das}},\
  and\ \bibinfo {author} {\bibfnamefont {R.}~\bibnamefont {Moessner}},\ }\href
  {https://doi.org/10.1103/PhysRevLett.115.030402} {\bibfield  {journal}
  {\bibinfo  {journal} {Phys. Rev. Lett.}\ }\textbf {\bibinfo {volume} {115}},\
  \bibinfo {pages} {030402} (\bibinfo {year} {2015})}\BibitemShut {NoStop}%
\bibitem [{\citenamefont {Bukov}\ \emph {et~al.}(2015)\citenamefont {Bukov},
  \citenamefont {D'Alessio},\ and\ \citenamefont {Polkovnikov}}]{Bukov2015}%
  \BibitemOpen
  \bibfield  {author} {\bibinfo {author} {\bibfnamefont {M.}~\bibnamefont
  {Bukov}}, \bibinfo {author} {\bibfnamefont {L.}~\bibnamefont {D'Alessio}},\
  and\ \bibinfo {author} {\bibfnamefont {A.}~\bibnamefont {Polkovnikov}},\
  }\href {https://doi.org/10.1080/00018732.2015.1055918} {\bibfield  {journal}
  {\bibinfo  {journal} {Advances in Physics}\ }\textbf {\bibinfo {volume}
  {64}},\ \bibinfo {pages} {139} (\bibinfo {year} {2015})}\BibitemShut
  {NoStop}%
\bibitem [{\citenamefont {Abanin}\ \emph {et~al.}(2016)\citenamefont {Abanin},
  \citenamefont {{De Roeck}},\ and\ \citenamefont {Huveneers}}]{Abanin2016}%
  \BibitemOpen
  \bibfield  {author} {\bibinfo {author} {\bibfnamefont {D.~A.}\ \bibnamefont
  {Abanin}}, \bibinfo {author} {\bibfnamefont {W.}~\bibnamefont {{De Roeck}}},\
  and\ \bibinfo {author} {\bibfnamefont {F.}~\bibnamefont {Huveneers}},\ }\href
  {https://doi.org/10.1016/j.aop.2016.03.010} {\bibfield
  {journal} {\bibinfo  {journal} {Annals of Physics}\ }\textbf {\bibinfo
  {volume} {372}},\ \bibinfo {pages} {1} (\bibinfo {year} {2016})}\BibitemShut
  {NoStop}%
\bibitem [{\citenamefont {Mori}\ \emph {et~al.}(2018)\citenamefont {Mori},
  \citenamefont {Ikeda}, \citenamefont {Kaminishi},\ and\ \citenamefont
  {Ueda}}]{Mori2018}%
  \BibitemOpen
  \bibfield  {author} {\bibinfo {author} {\bibfnamefont {T.}~\bibnamefont
  {Mori}}, \bibinfo {author} {\bibfnamefont {T.~N.}\ \bibnamefont {Ikeda}},
  \bibinfo {author} {\bibfnamefont {E.}~\bibnamefont {Kaminishi}},\ and\
  \bibinfo {author} {\bibfnamefont {M.}~\bibnamefont {Ueda}},\ }\href
  {https://doi.org/10.1088/1361-6455/aabcdf} {\bibfield  {journal} {\bibinfo
  {journal} {Journal of Physics B: Atomic, Molecular and Optical Physics}\
  }\textbf {\bibinfo {volume} {51}},\ \bibinfo {pages} {112001} (\bibinfo
  {year} {2018})}\BibitemShut {NoStop}%
\bibitem [{\citenamefont {Doggen}\ \emph {et~al.}(2021)\citenamefont {Doggen},
  \citenamefont {Gornyi}, \citenamefont {Mirlin},\ and\ \citenamefont
  {Polyakov}}]{Doggen2021}%
  \BibitemOpen
  \bibfield  {author} {\bibinfo {author} {\bibfnamefont {E.~V.}\ \bibnamefont
  {Doggen}}, \bibinfo {author} {\bibfnamefont {I.~V.}\ \bibnamefont {Gornyi}},
  \bibinfo {author} {\bibfnamefont {A.~D.}\ \bibnamefont {Mirlin}},\ and\
  \bibinfo {author} {\bibfnamefont {D.~G.}\ \bibnamefont {Polyakov}},\ }\href
  {https://doi.org/10.1016/j.aop.2021.168437} {\bibfield
  {journal} {\bibinfo  {journal} {Annals of Physics}\ }\textbf {\bibinfo
  {volume} {435}},\ \bibinfo {pages} {168437} (\bibinfo {year} {2021})},\
  \bibinfo {note} {special Issue on Localisation 2020}\BibitemShut {NoStop}%
\bibitem [{\citenamefont {Sierant}\ \emph {et~al.}(2023)\citenamefont
  {Sierant}, \citenamefont {Lewenstein}, \citenamefont {Scardicchio},\ and\
  \citenamefont {Zakrzewski}}]{PhysRevB.107.115132}%
  \BibitemOpen
  \bibfield  {author} {\bibinfo {author} {\bibfnamefont {P.}~\bibnamefont
  {Sierant}}, \bibinfo {author} {\bibfnamefont {M.}~\bibnamefont {Lewenstein}},
  \bibinfo {author} {\bibfnamefont {A.}~\bibnamefont {Scardicchio}},\ and\
  \bibinfo {author} {\bibfnamefont {J.}~\bibnamefont {Zakrzewski}},\ }\href
  {https://doi.org/10.1103/PhysRevB.107.115132} {\bibfield  {journal} {\bibinfo
   {journal} {Phys. Rev. B}\ }\textbf {\bibinfo {volume} {107}},\ \bibinfo
  {pages} {115132} (\bibinfo {year} {2023})}\BibitemShut {NoStop}%
\bibitem [{\citenamefont {Fletcher}\ \emph {et~al.}(2007)\citenamefont
  {Fletcher}, \citenamefont {Shor},\ and\ \citenamefont
  {Win}}]{FletcherShorWin2007}%
  \BibitemOpen
  \bibfield  {author} {\bibinfo {author} {\bibfnamefont {A.~S.}\ \bibnamefont
  {Fletcher}}, \bibinfo {author} {\bibfnamefont {P.~W.}\ \bibnamefont {Shor}},\
  and\ \bibinfo {author} {\bibfnamefont {M.~Z.}\ \bibnamefont {Win}},\ }\href
  {https://doi.org/10.1103/PhysRevA.75.012338} {\bibfield  {journal} {\bibinfo
  {journal} {Physical Review A}\ }\textbf {\bibinfo {volume} {75}},\ \bibinfo
  {pages} {012338} (\bibinfo {year} {2007})}\BibitemShut {NoStop}%
\bibitem [{\citenamefont {Prosen}(2000)}]{Prosen2000}%
  \BibitemOpen
  \bibfield  {author} {\bibinfo {author} {\bibfnamefont {T.}~\bibnamefont
  {Prosen}},\ }\href {https://doi.org/10.1143/PTPS.139.191} {\bibfield
  {journal} {\bibinfo  {journal} {Progress of Theoretical Physics Supplement}\
  }\textbf {\bibinfo {volume} {139}},\ \bibinfo {pages} {191} (\bibinfo {year}
  {2000})}\BibitemShut {NoStop}%
\bibitem [{\citenamefont {\ifmmode~\check{S}\else \v{S}\fi{}untajs}\ \emph
  {et~al.}(2020)\citenamefont {\ifmmode~\check{S}\else \v{S}\fi{}untajs},
  \citenamefont {Bon\ifmmode~\check{c}\else \v{c}\fi{}a}, \citenamefont
  {Prosen},\ and\ \citenamefont {Vidmar}}]{PhysRevE.102.062144}%
  \BibitemOpen
  \bibfield  {author} {\bibinfo {author} {\bibfnamefont {J.}~\bibnamefont
  {\ifmmode~\check{S}\else \v{S}\fi{}untajs}}, \bibinfo {author} {\bibfnamefont
  {J.}~\bibnamefont {Bon\ifmmode~\check{c}\else \v{c}\fi{}a}}, \bibinfo
  {author} {\bibfnamefont {T.~c.~v.}\ \bibnamefont {Prosen}},\ and\ \bibinfo
  {author} {\bibfnamefont {L.}~\bibnamefont {Vidmar}},\ }\href
  {https://doi.org/10.1103/PhysRevE.102.062144} {\bibfield  {journal} {\bibinfo
   {journal} {Phys. Rev. E}\ }\textbf {\bibinfo {volume} {102}},\ \bibinfo
  {pages} {062144} (\bibinfo {year} {2020})}\BibitemShut {NoStop}%
\bibitem [{\citenamefont {De~Roeck}\ and\ \citenamefont
  {Huveneers}(2017)}]{PhysRevB.95.155129}%
  \BibitemOpen
  \bibfield  {author} {\bibinfo {author} {\bibfnamefont {W.}~\bibnamefont
  {De~Roeck}}\ and\ \bibinfo {author} {\bibfnamefont {F.~m.~c.}\ \bibnamefont
  {Huveneers}},\ }\href {https://doi.org/10.1103/PhysRevB.95.155129} {\bibfield
   {journal} {\bibinfo  {journal} {Phys. Rev. B}\ }\textbf {\bibinfo {volume}
  {95}},\ \bibinfo {pages} {155129} (\bibinfo {year} {2017})}\BibitemShut
  {NoStop}%
\bibitem [{\citenamefont {Gopalakrishnan}\ and\ \citenamefont
  {Parameswaran}(2020)}]{Gopalakrishnan2020}%
  \BibitemOpen
  \bibfield  {author} {\bibinfo {author} {\bibfnamefont {S.}~\bibnamefont
  {Gopalakrishnan}}\ and\ \bibinfo {author} {\bibfnamefont {S.}~\bibnamefont
  {Parameswaran}},\ }\href
  {https://doi.org/10.1016/j.physrep.2020.03.003} {\bibfield
  {journal} {\bibinfo  {journal} {Physics Reports}\ }\textbf {\bibinfo {volume}
  {862}},\ \bibinfo {pages} {1} (\bibinfo {year} {2020})},\ \bibinfo {note}
  {dynamics and transport at the threshold of many-body
  localization}\BibitemShut {NoStop}%
\bibitem [{\citenamefont {Serbyn}\ \emph {et~al.}(2013)\citenamefont {Serbyn},
  \citenamefont {Papi\ifmmode~\acute{c}\else \'{c}\fi{}},\ and\ \citenamefont
  {Abanin}}]{PhysRevLett.111.127201}%
  \BibitemOpen
  \bibfield  {author} {\bibinfo {author} {\bibfnamefont {M.}~\bibnamefont
  {Serbyn}}, \bibinfo {author} {\bibfnamefont {Z.}~\bibnamefont
  {Papi\ifmmode~\acute{c}\else \'{c}\fi{}}},\ and\ \bibinfo {author}
  {\bibfnamefont {D.~A.}\ \bibnamefont {Abanin}},\ }\href
  {https://doi.org/10.1103/PhysRevLett.111.127201} {\bibfield  {journal}
  {\bibinfo  {journal} {Phys. Rev. Lett.}\ }\textbf {\bibinfo {volume} {111}},\
  \bibinfo {pages} {127201} (\bibinfo {year} {2013})}\BibitemShut {NoStop}%
\bibitem [{\citenamefont {Huse}\ \emph {et~al.}(2014)\citenamefont {Huse},
  \citenamefont {Nandkishore},\ and\ \citenamefont {Oganesyan}}]{Huse2014}%
  \BibitemOpen
  \bibfield  {author} {\bibinfo {author} {\bibfnamefont {D.~A.}\ \bibnamefont
  {Huse}}, \bibinfo {author} {\bibfnamefont {R.}~\bibnamefont {Nandkishore}},\
  and\ \bibinfo {author} {\bibfnamefont {V.}~\bibnamefont {Oganesyan}},\ }\href
  {https://doi.org/10.1103/PhysRevB.90.174202} {\bibfield  {journal} {\bibinfo
  {journal} {Physical Review B}\ }\textbf {\bibinfo {volume} {90}},\ \bibinfo
  {pages} {174202} (\bibinfo {year} {2014})}\BibitemShut {NoStop}%
\bibitem [{\citenamefont {Imbrie}\ \emph {et~al.}(2017)\citenamefont {Imbrie},
  \citenamefont {Ros},\ and\ \citenamefont
  {Scardicchio}}]{ImbrieRosScardicchio2017}%
  \BibitemOpen
  \bibfield  {author} {\bibinfo {author} {\bibfnamefont {J.~Z.}\ \bibnamefont
  {Imbrie}}, \bibinfo {author} {\bibfnamefont {V.}~\bibnamefont {Ros}},\ and\
  \bibinfo {author} {\bibfnamefont {A.}~\bibnamefont {Scardicchio}},\ }\href
  {https://doi.org/10.1002/andp.201600278} {\bibfield
  {journal} {\bibinfo  {journal} {Annalen der Physik}\ }\textbf {\bibinfo
  {volume} {529}},\ \bibinfo {pages} {1600278} (\bibinfo {year}
  {2017})}\BibitemShut {NoStop}%
\bibitem [{\citenamefont {Bauer}\ and\ \citenamefont
  {Nayak}(2013)}]{Bauer2013}%
  \BibitemOpen
  \bibfield  {author} {\bibinfo {author} {\bibfnamefont {B.}~\bibnamefont
  {Bauer}}\ and\ \bibinfo {author} {\bibfnamefont {C.}~\bibnamefont {Nayak}},\
  }\href {https://doi.org/10.1088/1742-5468/2013/09/P09005} {\bibfield
  {journal} {\bibinfo  {journal} {Journal of Statistical Mechanics: Theory and
  Experiment}\ }\textbf {\bibinfo {volume} {2013}},\ \bibinfo {pages} {P09005}
  (\bibinfo {year} {2013})}\BibitemShut {NoStop}%
\bibitem [{\citenamefont {Buijsman}\ \emph {et~al.}(2018)\citenamefont
  {Buijsman}, \citenamefont {Gritsev},\ and\ \citenamefont
  {Cheianov}}]{BuijsmanGritsevCheianov2018}%
  \BibitemOpen
  \bibfield  {author} {\bibinfo {author} {\bibfnamefont {W.}~\bibnamefont
  {Buijsman}}, \bibinfo {author} {\bibfnamefont {V.}~\bibnamefont {Gritsev}},\
  and\ \bibinfo {author} {\bibfnamefont {V.}~\bibnamefont {Cheianov}},\ }\href
  {https://doi.org/10.21468/SciPostPhys.4.6.038} {\bibfield  {journal}
  {\bibinfo  {journal} {SciPost Phys.}\ }\textbf {\bibinfo {volume} {4}},\
  \bibinfo {pages} {038} (\bibinfo {year} {2018})}\BibitemShut {NoStop}%
\bibitem [{\citenamefont {Nation}\ \emph {et~al.}(2021)\citenamefont {Nation},
  \citenamefont {Kang}, \citenamefont {Sundaresan},\ and\ \citenamefont
  {Gambetta}}]{PRXQuantum.2.040326}%
  \BibitemOpen
  \bibfield  {author} {\bibinfo {author} {\bibfnamefont {P.~D.}\ \bibnamefont
  {Nation}}, \bibinfo {author} {\bibfnamefont {H.}~\bibnamefont {Kang}},
  \bibinfo {author} {\bibfnamefont {N.}~\bibnamefont {Sundaresan}},\ and\
  \bibinfo {author} {\bibfnamefont {J.~M.}\ \bibnamefont {Gambetta}},\ }\href
  {https://doi.org/10.1103/PRXQuantum.2.040326} {\bibfield  {journal} {\bibinfo
   {journal} {PRX Quantum}\ }\textbf {\bibinfo {volume} {2}},\ \bibinfo {pages}
  {040326} (\bibinfo {year} {2021})}\BibitemShut {NoStop}%
\bibitem [{\citenamefont {Wu}\ \emph {et~al.}(2025)\citenamefont {Wu},
  \citenamefont {An}, \citenamefont {Zhang}, \citenamefont {Zhu},
  \citenamefont {Huang},\ and\ \citenamefont
  {Zeng}}]{WuAnZhangZhuHuangZeng2025}%
  \BibitemOpen
  \bibfield  {author} {\bibinfo {author} {\bibfnamefont {J.}~\bibnamefont
  {Wu}}, \bibinfo {author} {\bibfnamefont {Z.}~\bibnamefont {An}}, \bibinfo
  {author} {\bibfnamefont {C.}~\bibnamefont {Zhang}}, \bibinfo {author}
  {\bibfnamefont {X.}~\bibnamefont {Zhu}}, \bibinfo {author} {\bibfnamefont
  {S.}~\bibnamefont {Huang}},\ and\ \bibinfo {author} {\bibfnamefont {B.}~\bibnamefont
  {Zeng}},\ }\href {https://doi.org/10.1103/p7kb-k52j} {\bibfield  {journal}
  {\bibinfo  {journal} {Phys. Rev. A}\ }\textbf {\bibinfo {volume} {112}},\
  \bibinfo {pages} {062421} (\bibinfo {year} {2025})}\BibitemShut {NoStop}%
\end{thebibliography}
\end{document}